\documentclass[reprint,superscriptaddress,amsmath,amsfonts,amssymb,aps,prl,showkeys, longbibliography]{revtex4-2}
\usepackage{hyperref}
\usepackage{graphicx}
\usepackage{xcolor}
\hypersetup{
  breaklinks = {true},
  citecolor = {blue},
  colorlinks = {true},
  linkcolor = {red},
}
\usepackage{soul}
\usepackage{lipsum}
\usepackage{bm}

\newcommand{\JMD}[1]{{#1}}

\begin{document}

\title{Topologically Driven Spin-Orbit Torque in Dirac Matter} 

\author{Joaqu{\'i}n Medina Due{\~n}as}
\affiliation{ICN2 --- Catalan Institute of Nanoscience and Nanotechnology, CSIC and BIST, Campus UAB, Bellaterra, 08193 Barcelona, Spain}
\affiliation{Department of Physics, Universitat Aut{\`o}noma de Barcelona (UAB), Campus UAB, Bellaterra, 08193 Barcelona, Spain}

\author{Jos{\'e} H. Garc{\'i}a}
\email[Corresponding Author ]{josehugo.garcia@icn2.cat}
\affiliation{ICN2 --- Catalan Institute of Nanoscience and Nanotechnology, CSIC and BIST, Campus UAB, Bellaterra, 08193 Barcelona, Spain}

\author{Stephan Roche}
\affiliation{ICN2 --- Catalan Institute of Nanoscience and Nanotechnology, CSIC and BIST,
Campus UAB, Bellaterra, 08193 Barcelona, Spain}
\affiliation{ICREA --- Instituci\'o Catalana de Recerca i Estudis Avan\c{c}ats, 08010 Barcelona, Spain}

\date{\today}

\begin{abstract}
We unveil novel spin-orbit torque mechanisms driven by topological edge states in magnetic graphene-based devices. Within the energy gap, a damping-like torque plateau emerges within the quantum anomalous Hall phase upon breaking particle-hole symmetry, while for energies at the spin-split Dirac points located within the bands, a large damping-like torque develops as a result of a vanishing Fermi contour. Such torques are tunable by the degree of spin-pseudospin entanglement dictated by proximity-induced spin-orbit coupling terms.
\end{abstract}

\maketitle

The spin-orbit torque (SOT) mechanism represents a novel method to electrically manipulate a magnetic state, providing power-efficiency and miniaturization prospects beyond traditional multi-ferromagnet setups \cite{manchon_current-induced_2019}.
While the traditional spin-transfer torque mechanism relies on the generation of spin-polarized currents using ferromagnets \cite{slonczewski_current-driven_1996}, the SOT mechanism leverages SOC within metals to generate a current-driven spin source via charge-to-spin conversion \cite{miron_current-driven_2010, manchon_description_2008}. The non-equilibrium spin source $\boldsymbol{S}$ couples to the magnetic state, oriented along $\hat{\boldsymbol{m}}$, exerting a torque $\boldsymbol{T} \propto \hat{\boldsymbol{m}} \times \boldsymbol{S}$, driving the magnetization dynamics to eventually flip the magnetic state.
The conventional SOT phenomenon focuses on heavy metal/ferromagnet bilayers, considering two contributions: the damping-like (DL) torque which stems from a spin current injected from the bulk metal to the ferromagnet via the spin Hall effect, and the field-like (FL) torque generated by an interfacial spin density via the Rashba-Edelstein effect (REE).

Research in alternative systems, such as van der Waals materials and single-layered devices, has revealed that the origin of SOTs could be actually more complex \cite{tian_two-dimensional_2021, kurebayashi_magnetism_2022}.
Despite the suppression of the perpendicular spin Hall effect, DL torques have been measured in single-layered devices \cite{kurebayashi_antidamping_2014, seki_spin-orbit_2021, aoki_gigantic_2023} and two-dimensional materials \cite{zhang_spin_2016, macneill_thickness_2017, husain_large_2020} and were originally explained by intrinsic Berry curvature effects \cite{kurebayashi_antidamping_2014}, while the role of non-equilibrium spin textures has recently been revealed \cite{medina-duenas_emerging_2024}.
The development of atomically smooth interfaces demonstrates the fundamental role of crystal symmetries for generating unconventional torque contributions, eventually achieving field-free perpendicular magnetization switching \cite{macneill_control_2017, shi_all-electric_2019, liu_symmetry-dependent_2021}; although the spin or orbital nature of such torque remains debated \cite{ye_orbit-transfer_2022}.
Furthermore, the interplay between spin and other forms of angular momentum has been shown to enable novel SOT effects \cite{medina-duenas_emerging_2024, canonico_spin-orbit_2023}, where the band structure topology might also play a role \cite{johansson_spin_2021}. 
In graphene, spin-orbit coupling (SOC) is mediated by the pseudospin, which corresponds to an additional angular momentum degree of freedom stemming from the lattice \cite{mecklenburg_spin_2011}, and spin-pseudospin correlations have been argued to play a central role in spin transport and charge-to-spin conversion \cite{tuan_pseudospin-driven_2014,offidani_optimal_2017,de-moraes_emergence_2020}. Indeed, it has been shown that low-energy spin relaxation in graphene is driven by spin-pseudospin locking \cite{tuan_pseudospin-driven_2014,de-moraes_emergence_2020}, while the REE is limited by the quenching of the spin texture due to spin-pseudospin entanglement \cite{offidani_optimal_2017}. More recently, spin-pseudospin entanglement was also found to drive SOT beyond the semi-classical spin dynamics for Rashba-like SOC fields \cite{medina-duenas_emerging_2024}.

In this Letter we unveil a novel SOT of topological origin generated by spin-pseudospin entanglement in graphene-based heterostructures where magnetic exchange coupling coexists with Rashba and Kane-Mele SOC fields. 
We first demonstrate the manipulation of spin-pseudospin entanglement and its potential for enhancing charge-to-spin conversion by tuning the interplay between different forms of SOC. By these means, charge-to-spin conversion with \textit{maximal efficiency} is achieved via the REE in non-magnetic graphene-based heterostructures. 
Upon including a magnetic exchange coupling we find novel SOT mechanisms which are enabled by tuning spin-pseudospin entanglement at band inversions. Remarkably, we observe a DL torque plateau within the band gap emerging in the insulating topological phase of the quantum anomalous Hall effect, which is activated upon breaking particle-hole symmetry. Additionally, at the spin-split Dirac points a strong DL torque is generated from a vanishing Fermi contour. 
We trace back the origin of these torques to band structure properties to understand their microscopic origin. Our results unveil novel SOT mechanisms driven by topological phases, as well as manifesting the central role of correlations between spin and other forms of angular momentum for developing optimal spintronic platforms.

\begin{figure}[t!]
    \centering
    \includegraphics[width=\columnwidth]{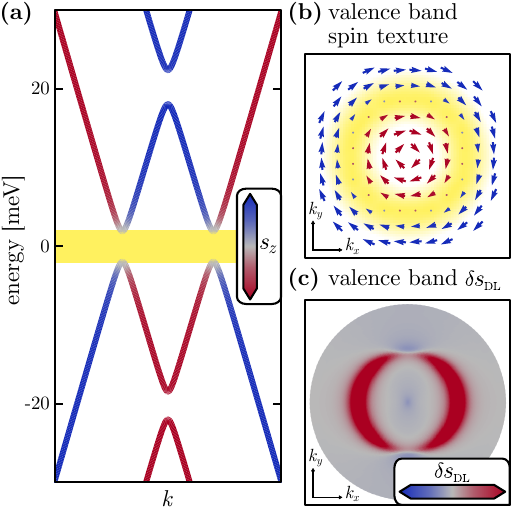}
    \caption{\textbf{(a)} Band structure exhibiting spin-split Dirac cones at energies $\mp J_\text{ex} / 2$ along the out-of-plane axis $\hat{\boldsymbol{z}}$, and a band gap at charge neutrality (yellow highlight), with $J_\text{ex} = 40 \,\text{meV}$ and $\lambda_\text{KM} = \lambda_\text{R} = 4 \,\text{meV}$ \textbf{(b)} Spin texture of the valence band, showing a sharp quenching at the band edge (yellow highlight) due to Kane-Mele SOC. \textbf{(c)} Damping-like non-equilibrium spin texture (arbitrary units) of the valence band, showing a source of DL torque at the band edge and at the Dirac point.}
    \label{fig:neqS}
\end{figure}

\JMD{
We consider a minimal graphene model comprising both Rashba and Kane-Mele SOC with an out-of-plane magnetization. The Hamiltonian is given by
\begin{equation}
\begin{split}
    \mathcal{H}_{\bm{k}} =&\, \hbar v (\tau k_x \sigma_x + k_y \sigma_y) - \frac{\lambda_\text{R}}{2} (s_x \sigma_y - \tau s_y \sigma_x) \\ &\, -\frac{\lambda_\text{KM}}{2} \tau s_z \sigma_z - \frac{J_\text{ex}}{2} s_z \text{ ,}
\end{split}
\end{equation}
with $v \sim 10^6 \,\text{m/s}$ the velocity of massless Dirac electrons in graphene, $\lambda_\text{R}$ and $\lambda_\text{KM}$ the Rashba and Kane-Mele SOC parameters respectively, and $J_\text{ex}$ the exchange splitting along the magnetization direction $\hat{\bm{z}}$. The spin and pseudospin degrees of freedom are respectively represented by the Pauli vectors $\bm{s}$ and $\bm{\sigma}$, while $\tau=\pm$ represents the valley index, where we will focus our analysis on the $\tau=+$ subspace without loss of generality \cite{supmat}. While SOC in bare graphene is negligible, such SOC fields can be induced by proximity effects in graphene/topological insulator \cite{khokhriakov_tailoring_2018, lee_proximity_2015, song_spin_2018, naimer_twist-angle_2023, jin_proximity-induced_2013} and graphene/transition metal dichalcogenide heterostructures \cite{wakamura_strong_2018, wakamura_spin-orbit_2019, avsar_spin-orbit_2014, wang_strong_2015, wang_origin_2016, gmitra_graphene_2015}. On the other hand, the magnetization term is obtained by interfacing graphene with an insulating ferromagnet \cite{tang_approaching_2017, wei_strong_2016, wu_magnetic_2017, song_assymetric_2018, wu_large_2020, ibrahim_unveiling_2019}.

The exchange splitting is usually the dominant energy scale, splitting the energy dispersion into spin majority and spin minority Dirac cones centered about energy $\mp J_\text{ex}/2$. Rashba SOC opens a gap at the crossing between both cones and induces a helical component to the spin texture, as shown in Fig.~\ref{fig:neqS}-(a).

The non-equilibrium spin density emerging from the REE, which is a Fermi-surface phenomenon, is computed using a Boltzmann approach as
\begin{equation}
    \boldsymbol{S}_\text{surf} (\varepsilon_\text{F}) = \sum_{n} \int \frac{\text{d}^2\boldsymbol{k}}{(2\pi)^2} \, \delta\mathit{f}_{n, \boldsymbol{k}} \boldsymbol{s}_{n,\boldsymbol{k}} \text{ ,}
\end{equation}
where $\varepsilon_\text{F}$ is the Fermi level, $\bm{s}_{n,\bm{k}} \equiv \langle E_{n,\bm{k}} | \bm{s} | E_{n,\bm{k}} \rangle$ is the spin texture, with $n$ the band index and $| E_{n,\bm{k}} \rangle$ the state of band $n$ at momentum $\bm{k}$, and $\delta\mathit{f}_{n,\bm{k}}$ is the current-induced variation of the carrier occupation. The resulting spin density is generated by the helical component of the spin texture which is anti-symmetric in $\bm{k}$, which under an applied electric field $\bm{\mathcal{E}}$ yields the a spin density of the form $S_\text{FL} \propto \hat{\bm{z}} \times \bm{\mathcal{E}}$, generating the FL torque.

Including the persistent out-of-plane spin texture generated by $J_\text{ex}$, symmetric in $\bm{k}$, and additional DL torque of the form $S_\text{DL} \propto m_z \bm{\mathcal{E}}$ emerges from the entire Fermi-sea contribution of non-equilibrium spin textures.
}
The Fermi sea non-equilibrium spin density is calculated as 
\begin{equation}
    \boldsymbol{S}_\text{sea} (\varepsilon_\text{F}) = \sum_n \int \frac{\text{d}^2 \boldsymbol{k}}{(2\pi)^2} \, \mathit{f}_{n,\boldsymbol{k}} \delta \boldsymbol{s}_{n,\boldsymbol{k}} \text{ ,}
\end{equation}
where $\mathit{f}_{n,\boldsymbol{k}}$ is the equilibrium carrier occupation, and $\delta \boldsymbol{s}_{n,\boldsymbol{k}}$ corresponds to a non-equilibrium component of the spin texture. 
We first describe this mechanism in the simple case of a 2D electron gas, as proposed in Ref.~\cite{medina-duenas_emerging_2024}. 
When applying an electric field, $\boldsymbol{\mathcal{E}}$, the wave vector of the Bloch electrons is modified as $\hbar \frac{\text{d}}{\text{d}t} \boldsymbol{k} = - e \boldsymbol{\mathcal{E}}$ (with $e$ the elementary charge), thus generating a non-equilibrium effective magnetic field stemming from spin-momentum locking in a moving rest-frame. The interaction between the equilibrium spin polarization and the non-equilibrium field yields a non-equilibrium component to the spin texture throughout the entire Fermi sea, captured by $\delta \boldsymbol{s}_{\boldsymbol{k}}$ \cite{medina-duenas_emerging_2024}.

In graphene this interpretation based on non-equilibrium spin dynamics is more complicated as the spin is not only coupled to a classical field, but also to the pseudospin angular momentum operator. Nevertheless, the semi-classical picture may be expanded to study the coupled spin-pseudospin dynamics given by the set of operators $\langle s_i\rangle$, $\langle \sigma_j \rangle$ and $\langle s_i \sigma_j \rangle$, with $i,j=x,y,z$; whose dynamics are determined by the Ehrenfest theorem. 
Strong spin-momentum locking serves as a landmark for a large DL torque, generated by the coupling between spin and an effective magnetic field. An additional DL torque landmark emerges from the coupling between spin and pseudospin, represented by spin-pseudospin entanglement hot spots at avoided band crossings.
In the supplementary material we develop our methodology in detail, showing that it yields equivalent results to the Kubo formalism, while providing a deeper grasp of the subjacent microscopic processes \cite{supmat}.

\begin{figure*}[t]
    \centering
    \includegraphics[width=\textwidth]{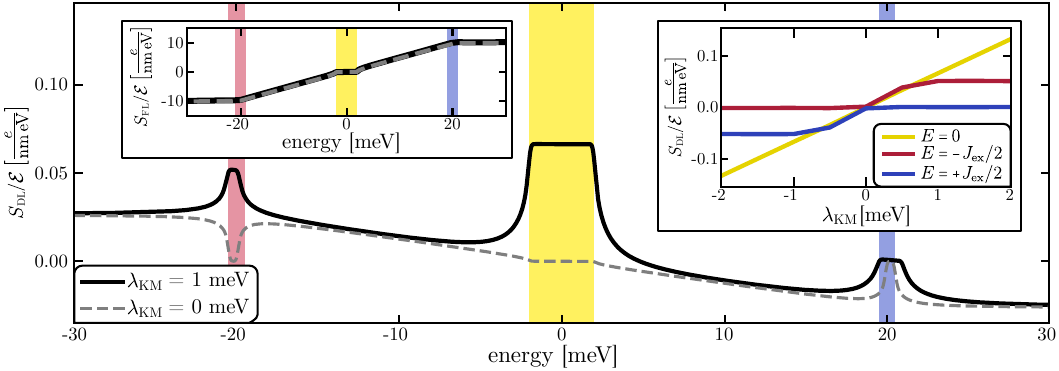}
    \caption{$S_\text{DL}$ with (solid black curve) and without (dashed grey curve) Kane-Mele SOC, with $J_\text{ex} = 40 \,\text{meV}$ and $\lambda_\text{R} = 4 \, \text{meV}$. A SOT plateau is observed throughout the band gap at charge neutrality enabled by $\lambda_\text{KM}$ (yellow highlight). At the Dirac points (red and blue highlights) a sharp valley in $S_\text{DL}$ is observed due to spin-pseudospin entanglement, which can be reverted into a sharp peak by $\lambda_\text{KM}$. Left inset: $S_\text{FL}$, is insensitive to $\lambda_\text{KM}$. Right inset: $S_\text{DL}$ at the gap (yellow), spin majority Dirac point (red), and spin minority Dirac cone (blue) as a function of $\lambda_\text{KM}$.}
    \label{fig:SOT}
\end{figure*}

The results for the non-equilibrium spin density, presented in Fig.~\ref{fig:SOT}, reveal that the DL spin density, $S_\text{DL}$, presents prominent features at the avoided band crossings, both at charge neutrality (yellow highlight), and at the spin-split Dirac points (red and blue highlights). Most remarkably, a SOT plateau is observed throughout the entire band gap at charge neutrality, where we additionally observe a quantized Hall conductivity of $\sigma_{xy} = 2e^2/h$ corresponding to a topological quantum anomalous Hall phase, consistent with previous studies in magnetic Rashba graphene \cite{qiao_quantum_2010, qiao_quantum_2014}. Interestingly however, while the quantum anomalous Hall phase remains quantized regardless of $\lambda_\text{KM}$, the $S_\text{DL}$ plateau only forms upon breaking particle-hole symmetry by including Kane-Mele SOC, and vanishes for $\lambda_\text{KM} = 0$ (see right inset of Fig. \ref{fig:SOT}). Thus, in absence of conductive bulk states, we observe the emergence of a DL torque which must be carried by the topological edge states related to the quantum anomalous Hall regime. Both of these quantities result from a non-equilibrium modification of the Bloch states throughout the Fermi sea due to adiabatic transport in $k$-space; however, while the Hall conductivity stems from a transverse charge current \cite{xiao_berry_2010}, we show here that $S_\text{DL}$ derives from a non-equilibrium spin texture. This behavior reveals a SOT driven by topological physics which has been so far unnoticed.

We offer an explanation for the origin of the SOT plateau based on the semi-classical spin-pseudospin dynamics, relating it to band structure features. As shown in Fig.~\ref{fig:neqS}-(a), the band gap at charge neutrality (yellow highlight) is generated by the avoided crossing between the upper band of the spin majority Dirac cone and the lower band of the spin minority one, with positive and negative out-of-plane spin polarization respectively and both exhibiting the same helicity. 
For $\lambda_\text{KM}=0$ these bands present a radial pseudospin polarization along $\pm \hat{\boldsymbol{k}}$, opposite to one another, belonging to orthogonal pseudospin subspaces. Thus, the hybridization between both bands near the gap leads to a loss of their out-of-plane spin polarization due to spin-pseudospin entanglement, while their in-plane spin texture is unaffected.
On the other hand, when introducing Kane-Mele SOC both bands acquire the same out-of-plane pseudospin component proportional to $\lambda_\text{KM}$, and thus share a common pseudospin subspace. In this case the hybridization between bands of up and down spin polarization stemming from the opposite Dirac cones modifies the in-plane spin texture \cite{supmat}. This process results in a sharp variation of the spin helicity at the band edges, as highlighted in yellow in Fig.~\ref{fig:neqS}-(b), where the mark of strong spin-momentum locking forecasts a large $S_\text{DL}$.
Indeed, as the Bloch states are transported under an applied electric field, those near the band edge feel a strong variation of the effective magnetic field, which drives the non-equilibrium spin dynamics resulting in a non-equilibrium spin texture source at the band edge, shown in Fig.~\ref{fig:neqS}-(c).
This effect is opposite at the valence and conduction bands, manifesting broken particle-hole symmetry, and therefore, within the gap the fully filled valence band generates the DL SOT plateau \cite{supmat}.

The FL torque, on the other hand, stems from the Fermi-surface REE and remains practically insensitive to $\lambda_\text{KM}$, as shown in the left inset of Fig.~\ref{fig:SOT}. Even though the spin helicity is modified by $\lambda_\text{KM}$ near the gap, $S_\text{FL}$ remains practically unchanged as the two Fermi contours of opposite velocities cancel one another. At higher energies the inner contour shrinks until vanishing when reaching the Dirac point, where $S_\text{FL}$ reaches its maximum value as the band stemming from the opposite Dirac cone exclusively dominates Fermi-surface transport. Kane-Mele SOC opens a semi-gap at the Dirac point; however, $S_\text{FL}$ is unaffected as it is solely determined by the outer Fermi contour.

On the contrary, the DL torque shows a remarkable behavior at the Dirac points, where a large non-equilibrium spin density is generated within the Kane-Mele semi-gap (shown by the red and blue highlights in Fig.~\ref{fig:SOT}), along with the corresponding spin Hall effect driven by the Berry curvature \cite{supmat}. Distinct to the SOT plateau at charge neutrality whose underlying mechanism relies on the enhancement of spin-momentum locking at the band edges, the spin density generated at the Dirac point stems from spin-pseudospin entanglement sources at the avoided band crossing. Rashba SOC not only generates helical spin-momentum locking, but also promotes spin-pseudospin entanglement by hybridizing the opposite Dirac cones. At $k=0$ the latter effect becomes more relevant due to the vanishing momentum, where, considering $\lambda_\text{KM} = 0$, the bottom-most and top-most bands hybridize to form spin-pseudospin entangled states. Thus, while at a finite momentum the two bands stemming from a same Dirac cone present opposite pseudospin polarization, near $k=0$ they share a common pseudospin subspace. By these means, under an applied electric field inter-band transitions at the Dirac point generate a large non-equilibrium spin density of opposite sign to the one generated by the outer Fermi contour, resulting in sharp $S_\text{DL}$ valleys at both Dirac cones for $\lambda_\text{KM} = 0$, shown by the dashed curve in Fig. \ref{fig:SOT}.

The DL torque at the Dirac cones can be tuned by Kane-Mele SOC. Indeed, Rashba SOC promotes a hybridization between the spin-split Dirac cones, which are spectrally separated by $J_\text{ex}$ at $k=0$, imprinting an effective mass of $\approx \lambda_\text{R}^2 / J_\text{ex}$ on both cones. On the other hand, Kane-Mele SOC induces an effective mass of opposite sign in each cone, breaking particle-hole symmetry. For $\lambda_\text{KM} > 0$ both contributions to the effective mass collaborate at the spin minority Dirac cone, enhancing the $S_\text{DL}$ valley (see blue highlight in Fig.\ref{fig:SOT}). However, at the spin majority Dirac cone the sign of the band inversion induced by Rashba SOC is flipped by $\lambda_\text{KM}$. The $S_\text{DL}$ contribution stemming from within the Dirac point now interferes constructively with the one stemming from the outer Fermi contour, generating a pronounced peak in the non-equilibrium spin density (see red highlight in Fig.\ref{fig:SOT}).

In conclusion, the role of spin-pseudospin entanglement and presence of topological states have been shown to generate a novel type of spin-orbit torque driven by topological physics, and maximally optimize charge-to-spin conversion via the Rashba-Edelstein effect. We expect these conclusions to be general to systems where spin-momentum locking is mediated by additional degrees of freedom, as can also occur with orbital angular momentum \cite{go_theory_2020}. Our findings open new paths for efficient spintronic devices, as well as for fundamental research on novel topological phases.

\begin{acknowledgments}
The authors acknowledge funding from the project  I+D+i PID2022-138283NB-I00 funded by MICIU/AEI/10.13039/501100011033/ and “FEDER Una manera de hacer Europa”, FLAG-ERA project MNEMOSYN (PCI2021-122035-2A) funded by MICIU/AEI /10.13039/501100011033 and the European Union NextGenerationEU/PRTR, and ERC project AI4SPIN funded by the  European Union’s Horizon Europe research and innovation programme - European Research Council Executive Agency under grant agreement No 101078370-AI4SPIN.
This work is also supported by MICIU with European funds‐NextGenerationEU (PRTR‐C17.I1) and by and 2021 SGR 00997, funded by Generalitat de Catalunya.
J.M.D. acknowledges support from MICIU grant FPI PRE2021-097031. 
ICN2 is funded by the CERCA Programme/Generalitat de Catalunya and supported by the Severo Ochoa Centres of Excellence programme, Grant CEX2021-001214-S, funded by MCIN/AEI/10.13039.501100011033.
 \end{acknowledgments}


\begin{thebibliography}{45}%
\makeatletter
\providecommand \@ifxundefined [1]{%
 \@ifx{#1\undefined}
}%
\providecommand \@ifnum [1]{%
 \ifnum #1\expandafter \@firstoftwo
 \else \expandafter \@secondoftwo
 \fi
}%
\providecommand \@ifx [1]{%
 \ifx #1\expandafter \@firstoftwo
 \else \expandafter \@secondoftwo
 \fi
}%
\providecommand \natexlab [1]{#1}%
\providecommand \enquote  [1]{``#1''}%
\providecommand \bibnamefont  [1]{#1}%
\providecommand \bibfnamefont [1]{#1}%
\providecommand \citenamefont [1]{#1}%
\providecommand \href@noop [0]{\@secondoftwo}%
\providecommand \href [0]{\begingroup \@sanitize@url \@href}%
\providecommand \@href[1]{\@@startlink{#1}\@@href}%
\providecommand \@@href[1]{\endgroup#1\@@endlink}%
\providecommand \@sanitize@url [0]{\catcode `\\12\catcode `\$12\catcode `\&12\catcode `\#12\catcode `\^12\catcode `\_12\catcode `\%12\relax}%
\providecommand \@@startlink[1]{}%
\providecommand \@@endlink[0]{}%
\providecommand \url  [0]{\begingroup\@sanitize@url \@url }%
\providecommand \@url [1]{\endgroup\@href {#1}{\urlprefix }}%
\providecommand \urlprefix  [0]{URL }%
\providecommand \Eprint [0]{\href }%
\providecommand \doibase [0]{https://doi.org/}%
\providecommand \selectlanguage [0]{\@gobble}%
\providecommand \bibinfo  [0]{\@secondoftwo}%
\providecommand \bibfield  [0]{\@secondoftwo}%
\providecommand \translation [1]{[#1]}%
\providecommand \BibitemOpen [0]{}%
\providecommand \bibitemStop [0]{}%
\providecommand \bibitemNoStop [0]{.\EOS\space}%
\providecommand \EOS [0]{\spacefactor3000\relax}%
\providecommand \BibitemShut  [1]{\csname bibitem#1\endcsname}%
\let\auto@bib@innerbib\@empty
\bibitem [{\citenamefont {Manchon}\ \emph {et~al.}(2019)\citenamefont {Manchon}, \citenamefont {{\v Z}elezn{\'y}}, \citenamefont {Miron}, \citenamefont {Jungwirth}, \citenamefont {Sinova}, \citenamefont {Thiaville}, \citenamefont {Garello},\ and\ \citenamefont {Gambardella}}]{manchon_current-induced_2019}%
  \BibitemOpen
  \bibfield  {author} {\bibinfo {author} {\bibfnamefont {A.}~\bibnamefont {Manchon}}, \bibinfo {author} {\bibfnamefont {J.}~\bibnamefont {{\v Z}elezn{\'y}}}, \bibinfo {author} {\bibfnamefont {I.~M.}\ \bibnamefont {Miron}}, \bibinfo {author} {\bibfnamefont {T.}~\bibnamefont {Jungwirth}}, \bibinfo {author} {\bibfnamefont {J.}~\bibnamefont {Sinova}}, \bibinfo {author} {\bibfnamefont {A.}~\bibnamefont {Thiaville}}, \bibinfo {author} {\bibfnamefont {K.}~\bibnamefont {Garello}},\ and\ \bibinfo {author} {\bibfnamefont {P.}~\bibnamefont {Gambardella}},\ }\bibfield  {title} {\bibinfo {title} {Current-induced spin-orbit torques in ferromagnetic and antiferromagnetic systems},\ }\href {https://doi.org/10.1103/RevModPhys.91.035004} {\bibfield  {journal} {\bibinfo  {journal} {Rev. Mod. Phys.}\ }\textbf {\bibinfo {volume} {91}},\ \bibinfo {pages} {035004} (\bibinfo {year} {2019})}\BibitemShut {NoStop}%
\bibitem [{\citenamefont {Slonczewski}(1996)}]{slonczewski_current-driven_1996}%
  \BibitemOpen
  \bibfield  {author} {\bibinfo {author} {\bibfnamefont {J.}~\bibnamefont {Slonczewski}},\ }\bibfield  {title} {\bibinfo {title} {Current-driven excitation of magnetic multilayers},\ }\href {https://doi.org/10.1016/0304-8853(96)00062-5} {\bibfield  {journal} {\bibinfo  {journal} {Journal of Magnetism and Magnetic Materials}\ }\textbf {\bibinfo {volume} {159}},\ \bibinfo {pages} {L1} (\bibinfo {year} {1996})}\BibitemShut {NoStop}%
\bibitem [{\citenamefont {Miron}\ \emph {et~al.}(2010)\citenamefont {Miron}, \citenamefont {Gaudin}, \citenamefont {Auffret}, \citenamefont {Rodmacq}, \citenamefont {Schuhl}, \citenamefont {Pizzini}, \citenamefont {Vogel},\ and\ \citenamefont {Gambardella}}]{miron_current-driven_2010}%
  \BibitemOpen
  \bibfield  {author} {\bibinfo {author} {\bibfnamefont {I.~M.}\ \bibnamefont {Miron}}, \bibinfo {author} {\bibfnamefont {G.}~\bibnamefont {Gaudin}}, \bibinfo {author} {\bibfnamefont {S.}~\bibnamefont {Auffret}}, \bibinfo {author} {\bibfnamefont {B.}~\bibnamefont {Rodmacq}}, \bibinfo {author} {\bibfnamefont {A.}~\bibnamefont {Schuhl}}, \bibinfo {author} {\bibfnamefont {S.}~\bibnamefont {Pizzini}}, \bibinfo {author} {\bibfnamefont {J.}~\bibnamefont {Vogel}},\ and\ \bibinfo {author} {\bibfnamefont {P.}~\bibnamefont {Gambardella}},\ }\bibfield  {title} {\bibinfo {title} {Current-driven spin torque induced by the {Rashba} effect in a ferromagnetic metal layer},\ }\href {https://doi.org/10.1038/nmat2613} {\bibfield  {journal} {\bibinfo  {journal} {Nature Mater}\ }\textbf {\bibinfo {volume} {9}},\ \bibinfo {pages} {230} (\bibinfo {year} {2010})}\BibitemShut {NoStop}%
\bibitem [{\citenamefont {Manchon}\ \emph {et~al.}(2008)\citenamefont {Manchon}, \citenamefont {Ryzhanova}, \citenamefont {Vedyayev}, \citenamefont {Chschiev},\ and\ \citenamefont {Dieny}}]{manchon_description_2008}%
  \BibitemOpen
  \bibfield  {author} {\bibinfo {author} {\bibfnamefont {A.}~\bibnamefont {Manchon}}, \bibinfo {author} {\bibfnamefont {N.}~\bibnamefont {Ryzhanova}}, \bibinfo {author} {\bibfnamefont {A.}~\bibnamefont {Vedyayev}}, \bibinfo {author} {\bibfnamefont {M.}~\bibnamefont {Chschiev}},\ and\ \bibinfo {author} {\bibfnamefont {B.}~\bibnamefont {Dieny}},\ }\bibfield  {title} {\bibinfo {title} {Description of current-driven torques in magnetic tunnel junctions},\ }\href {https://doi.org/10.1088/0953-8984/20/14/145208} {\bibfield  {journal} {\bibinfo  {journal} {Journal of Physics: Condensed Matter}\ }\textbf {\bibinfo {volume} {20}},\ \bibinfo {pages} {145208} (\bibinfo {year} {2008})}\BibitemShut {NoStop}%
\bibitem [{\citenamefont {Tian}\ \emph {et~al.}(2021)\citenamefont {Tian}, \citenamefont {Zhu}, \citenamefont {Jalali}, \citenamefont {Jiang}, \citenamefont {Liang}, \citenamefont {Huang}, \citenamefont {Chen}, \citenamefont {Zeng},\ and\ \citenamefont {Zhai}}]{tian_two-dimensional_2021}%
  \BibitemOpen
  \bibfield  {author} {\bibinfo {author} {\bibfnamefont {M.}~\bibnamefont {Tian}}, \bibinfo {author} {\bibfnamefont {Y.}~\bibnamefont {Zhu}}, \bibinfo {author} {\bibfnamefont {M.}~\bibnamefont {Jalali}}, \bibinfo {author} {\bibfnamefont {W.}~\bibnamefont {Jiang}}, \bibinfo {author} {\bibfnamefont {J.}~\bibnamefont {Liang}}, \bibinfo {author} {\bibfnamefont {Z.}~\bibnamefont {Huang}}, \bibinfo {author} {\bibfnamefont {Q.}~\bibnamefont {Chen}}, \bibinfo {author} {\bibfnamefont {Z.}~\bibnamefont {Zeng}},\ and\ \bibinfo {author} {\bibfnamefont {Y.}~\bibnamefont {Zhai}},\ }\bibfield  {title} {\bibinfo {title} {Two-dimensional van der waals materials for spin-orbit torque applications},\ }\bibfield  {journal} {\bibinfo  {journal} {Frontiers in Nanotechnology}\ }\textbf {\bibinfo {volume} {3}},\ \href {https://doi.org/10.3389/fnano.2021.732916} {10.3389/fnano.2021.732916} (\bibinfo {year} {2021})\BibitemShut {NoStop}%
\bibitem [{\citenamefont {Kurebayashi}\ \emph {et~al.}(2022)\citenamefont {Kurebayashi}, \citenamefont {Garc{\'i}a}, \citenamefont {Khan}, \citenamefont {Sinova},\ and\ \citenamefont {Roche}}]{kurebayashi_magnetism_2022}%
  \BibitemOpen
  \bibfield  {author} {\bibinfo {author} {\bibfnamefont {H.}~\bibnamefont {Kurebayashi}}, \bibinfo {author} {\bibfnamefont {J.~H.}\ \bibnamefont {Garc{\'i}a}}, \bibinfo {author} {\bibfnamefont {S.}~\bibnamefont {Khan}}, \bibinfo {author} {\bibfnamefont {J.}~\bibnamefont {Sinova}},\ and\ \bibinfo {author} {\bibfnamefont {S.}~\bibnamefont {Roche}},\ }\bibfield  {title} {\bibinfo {title} {Magnetism, symmetry and spin transport in van der {Waals} layered systems},\ }\href {https://doi.org/10.1038/s42254-021-00403-5} {\bibfield  {journal} {\bibinfo  {journal} {Nat. Rev. Phys.}\ }\textbf {\bibinfo {volume} {4}},\ \bibinfo {pages} {150} (\bibinfo {year} {2022})}\BibitemShut {NoStop}%
\bibitem [{\citenamefont {Kurebayashi}\ \emph {et~al.}(2014)\citenamefont {Kurebayashi}, \citenamefont {Sinova}, \citenamefont {Fang}, \citenamefont {Irvine}, \citenamefont {Skinner}, \citenamefont {Wunderlich}, \citenamefont {Nov{\'a}k}, \citenamefont {Campion}, \citenamefont {Gallagher}, \citenamefont {Vehstedt}, \citenamefont {Z{\^a}rbo}, \citenamefont {V{\'y}born{\'y}}, \citenamefont {Ferguson},\ and\ \citenamefont {Jungwirth}}]{kurebayashi_antidamping_2014}%
  \BibitemOpen
  \bibfield  {author} {\bibinfo {author} {\bibfnamefont {H.}~\bibnamefont {Kurebayashi}}, \bibinfo {author} {\bibfnamefont {J.}~\bibnamefont {Sinova}}, \bibinfo {author} {\bibfnamefont {D.}~\bibnamefont {Fang}}, \bibinfo {author} {\bibfnamefont {A.~C.}\ \bibnamefont {Irvine}}, \bibinfo {author} {\bibfnamefont {T.~D.}\ \bibnamefont {Skinner}}, \bibinfo {author} {\bibfnamefont {J.}~\bibnamefont {Wunderlich}}, \bibinfo {author} {\bibfnamefont {V.}~\bibnamefont {Nov{\'a}k}}, \bibinfo {author} {\bibfnamefont {R.~P.}\ \bibnamefont {Campion}}, \bibinfo {author} {\bibfnamefont {B.~L.}\ \bibnamefont {Gallagher}}, \bibinfo {author} {\bibfnamefont {E.~K.}\ \bibnamefont {Vehstedt}}, \bibinfo {author} {\bibfnamefont {L.~P.}\ \bibnamefont {Z{\^a}rbo}}, \bibinfo {author} {\bibfnamefont {K.}~\bibnamefont {V{\'y}born{\'y}}}, \bibinfo {author} {\bibfnamefont {A.~J.}\ \bibnamefont {Ferguson}},\ and\ \bibinfo {author} {\bibfnamefont {T.}~\bibnamefont {Jungwirth}},\ }\bibfield  {title} {\bibinfo {title} {An antidamping
  spin–orbit torque originating from the {Berry} curvature},\ }\href {https://doi.org/10.1038/nnano.2014.15} {\bibfield  {journal} {\bibinfo  {journal} {Nature Nanotechnol.}\ }\textbf {\bibinfo {volume} {9}},\ \bibinfo {pages} {211} (\bibinfo {year} {2014})}\BibitemShut {NoStop}%
\bibitem [{\citenamefont {Seki}\ \emph {et~al.}(2021)\citenamefont {Seki}, \citenamefont {Lau}, \citenamefont {Iihama},\ and\ \citenamefont {Takanashi}}]{seki_spin-orbit_2021}%
  \BibitemOpen
  \bibfield  {author} {\bibinfo {author} {\bibfnamefont {T.}~\bibnamefont {Seki}}, \bibinfo {author} {\bibfnamefont {Y.-C.}\ \bibnamefont {Lau}}, \bibinfo {author} {\bibfnamefont {S.}~\bibnamefont {Iihama}},\ and\ \bibinfo {author} {\bibfnamefont {K.}~\bibnamefont {Takanashi}},\ }\bibfield  {title} {\bibinfo {title} {Spin-orbit torque in a {Ni}-{Fe} single layer},\ }\href {https://doi.org/10.1103/PhysRevB.104.094430} {\bibfield  {journal} {\bibinfo  {journal} {Phys. Rev. B}\ }\textbf {\bibinfo {volume} {104}},\ \bibinfo {pages} {094430} (\bibinfo {year} {2021})}\BibitemShut {NoStop}%
\bibitem [{\citenamefont {Aoki}\ \emph {et~al.}(2023)\citenamefont {Aoki}, \citenamefont {Yin}, \citenamefont {Granville}, \citenamefont {Zhang}, \citenamefont {Medhekar}, \citenamefont {Leiva}, \citenamefont {Ohshima}, \citenamefont {Ando},\ and\ \citenamefont {Shiraishi}}]{aoki_gigantic_2023}%
  \BibitemOpen
  \bibfield  {author} {\bibinfo {author} {\bibfnamefont {M.}~\bibnamefont {Aoki}}, \bibinfo {author} {\bibfnamefont {Y.}~\bibnamefont {Yin}}, \bibinfo {author} {\bibfnamefont {S.}~\bibnamefont {Granville}}, \bibinfo {author} {\bibfnamefont {Y.}~\bibnamefont {Zhang}}, \bibinfo {author} {\bibfnamefont {N.~V.}\ \bibnamefont {Medhekar}}, \bibinfo {author} {\bibfnamefont {L.}~\bibnamefont {Leiva}}, \bibinfo {author} {\bibfnamefont {R.}~\bibnamefont {Ohshima}}, \bibinfo {author} {\bibfnamefont {Y.}~\bibnamefont {Ando}},\ and\ \bibinfo {author} {\bibfnamefont {M.}~\bibnamefont {Shiraishi}},\ }\bibfield  {title} {\bibinfo {title} {Gigantic anisotropy of self-induced spin-orbit torque in {Weyl} ferromagnet {Co2MnGa}},\ }\href {https://doi.org/10.1021/acs.nanolett.3c01573} {\bibfield  {journal} {\bibinfo  {journal} {Nano Letters}\ }\textbf {\bibinfo {volume} {23}},\ \bibinfo {pages} {6951} (\bibinfo {year} {2023})}\BibitemShut {NoStop}%
\bibitem [{\citenamefont {Zhang}\ \emph {et~al.}(2016)\citenamefont {Zhang}, \citenamefont {Sklenar}, \citenamefont {Hsu}, \citenamefont {Jiang}, \citenamefont {Jungfleisch}, \citenamefont {Xiao}, \citenamefont {Fradin}, \citenamefont {Liu}, \citenamefont {Pearson}, \citenamefont {Ketterson}, \citenamefont {Yang},\ and\ \citenamefont {Hoffmann}}]{zhang_spin_2016}%
  \BibitemOpen
  \bibfield  {author} {\bibinfo {author} {\bibfnamefont {W.}~\bibnamefont {Zhang}}, \bibinfo {author} {\bibfnamefont {J.}~\bibnamefont {Sklenar}}, \bibinfo {author} {\bibfnamefont {B.}~\bibnamefont {Hsu}}, \bibinfo {author} {\bibfnamefont {W.}~\bibnamefont {Jiang}}, \bibinfo {author} {\bibfnamefont {M.~B.}\ \bibnamefont {Jungfleisch}}, \bibinfo {author} {\bibfnamefont {J.}~\bibnamefont {Xiao}}, \bibinfo {author} {\bibfnamefont {F.~Y.}\ \bibnamefont {Fradin}}, \bibinfo {author} {\bibfnamefont {Y.}~\bibnamefont {Liu}}, \bibinfo {author} {\bibfnamefont {J.~E.}\ \bibnamefont {Pearson}}, \bibinfo {author} {\bibfnamefont {J.~B.}\ \bibnamefont {Ketterson}}, \bibinfo {author} {\bibfnamefont {Z.}~\bibnamefont {Yang}},\ and\ \bibinfo {author} {\bibfnamefont {A.}~\bibnamefont {Hoffmann}},\ }\bibfield  {title} {\bibinfo {title} {{Research Update: Spin transfer torques in permalloy on monolayer MoS2}},\ }\href {https://doi.org/10.1063/1.4943076} {\bibfield  {journal} {\bibinfo  {journal} {APL Materials}\ }\textbf
  {\bibinfo {volume} {4}},\ \bibinfo {pages} {032302} (\bibinfo {year} {2016})}\BibitemShut {NoStop}%
\bibitem [{\citenamefont {MacNeill}\ \emph {et~al.}(2017{\natexlab{a}})\citenamefont {MacNeill}, \citenamefont {Stiehl}, \citenamefont {Guimaraes}, \citenamefont {Reynolds}, \citenamefont {Buhrman},\ and\ \citenamefont {Ralph}}]{macneill_thickness_2017}%
  \BibitemOpen
  \bibfield  {author} {\bibinfo {author} {\bibfnamefont {D.}~\bibnamefont {MacNeill}}, \bibinfo {author} {\bibfnamefont {G.~M.}\ \bibnamefont {Stiehl}}, \bibinfo {author} {\bibfnamefont {M.~H.~D.}\ \bibnamefont {Guimaraes}}, \bibinfo {author} {\bibfnamefont {N.~D.}\ \bibnamefont {Reynolds}}, \bibinfo {author} {\bibfnamefont {R.~A.}\ \bibnamefont {Buhrman}},\ and\ \bibinfo {author} {\bibfnamefont {D.~C.}\ \bibnamefont {Ralph}},\ }\bibfield  {title} {\bibinfo {title} {Thickness dependence of spin-orbit torques generated by wte2},\ }\href {https://doi.org/10.1103/physrevb.96.054450} {\bibfield  {journal} {\bibinfo  {journal} {Physical Review B}\ }\textbf {\bibinfo {volume} {96}},\ \bibinfo {pages} {054450} (\bibinfo {year} {2017}{\natexlab{a}})}\BibitemShut {NoStop}%
\bibitem [{\citenamefont {Husain}\ \emph {et~al.}(2020)\citenamefont {Husain}, \citenamefont {Chen}, \citenamefont {Gupta}, \citenamefont {Behera}, \citenamefont {Kumar}, \citenamefont {Edvinsson}, \citenamefont {García-Sánchez}, \citenamefont {Brucas}, \citenamefont {Chaudhary}, \citenamefont {Sanyal}, \citenamefont {Svedlindh},\ and\ \citenamefont {Kumar}}]{husain_large_2020}%
  \BibitemOpen
  \bibfield  {author} {\bibinfo {author} {\bibfnamefont {S.}~\bibnamefont {Husain}}, \bibinfo {author} {\bibfnamefont {X.}~\bibnamefont {Chen}}, \bibinfo {author} {\bibfnamefont {R.}~\bibnamefont {Gupta}}, \bibinfo {author} {\bibfnamefont {N.}~\bibnamefont {Behera}}, \bibinfo {author} {\bibfnamefont {P.}~\bibnamefont {Kumar}}, \bibinfo {author} {\bibfnamefont {T.}~\bibnamefont {Edvinsson}}, \bibinfo {author} {\bibfnamefont {F.}~\bibnamefont {García-Sánchez}}, \bibinfo {author} {\bibfnamefont {R.}~\bibnamefont {Brucas}}, \bibinfo {author} {\bibfnamefont {S.}~\bibnamefont {Chaudhary}}, \bibinfo {author} {\bibfnamefont {B.}~\bibnamefont {Sanyal}}, \bibinfo {author} {\bibfnamefont {P.}~\bibnamefont {Svedlindh}},\ and\ \bibinfo {author} {\bibfnamefont {A.}~\bibnamefont {Kumar}},\ }\bibfield  {title} {\bibinfo {title} {Large damping-like spin–orbit torque in a 2d conductive 1t-tas2 monolayer},\ }\href {https://doi.org/10.1021/acs.nanolett.0c01955} {\bibfield  {journal} {\bibinfo  {journal} {Nano Letters}\
  }\textbf {\bibinfo {volume} {20}},\ \bibinfo {pages} {6372–6380} (\bibinfo {year} {2020})}\BibitemShut {NoStop}%
\bibitem [{\citenamefont {Medina Due\~nas}\ \emph {et~al.}(2024)\citenamefont {Medina Due\~nas}, \citenamefont {Garc\'{\i}a},\ and\ \citenamefont {Roche}}]{medina-duenas_emerging_2024}%
  \BibitemOpen
  \bibfield  {author} {\bibinfo {author} {\bibfnamefont {J.}~\bibnamefont {Medina Due\~nas}}, \bibinfo {author} {\bibfnamefont {J.~H.}\ \bibnamefont {Garc\'{\i}a}},\ and\ \bibinfo {author} {\bibfnamefont {S.}~\bibnamefont {Roche}},\ }\bibfield  {title} {\bibinfo {title} {Emerging spin-orbit torques in low-dimensional {Dirac} materials},\ }\href {https://doi.org/10.1103/PhysRevLett.132.266301} {\bibfield  {journal} {\bibinfo  {journal} {Phys. Rev. Lett.}\ }\textbf {\bibinfo {volume} {132}},\ \bibinfo {pages} {266301} (\bibinfo {year} {2024})}\BibitemShut {NoStop}%
\bibitem [{\citenamefont {MacNeill}\ \emph {et~al.}(2017{\natexlab{b}})\citenamefont {MacNeill}, \citenamefont {Stiehl}, \citenamefont {Guimaraes}, \citenamefont {Buhrman}, \citenamefont {Park},\ and\ \citenamefont {Ralph}}]{macneill_control_2017}%
  \BibitemOpen
  \bibfield  {author} {\bibinfo {author} {\bibfnamefont {D.}~\bibnamefont {MacNeill}}, \bibinfo {author} {\bibfnamefont {G.~M.}\ \bibnamefont {Stiehl}}, \bibinfo {author} {\bibfnamefont {M.~H.~D.}\ \bibnamefont {Guimaraes}}, \bibinfo {author} {\bibfnamefont {R.~A.}\ \bibnamefont {Buhrman}}, \bibinfo {author} {\bibfnamefont {J.}~\bibnamefont {Park}},\ and\ \bibinfo {author} {\bibfnamefont {D.~C.}\ \bibnamefont {Ralph}},\ }\bibfield  {title} {\bibinfo {title} {Control of spin–orbit torques through crystal symmetry in {WTe2}/ferromagnet bilayers},\ }\href {https://doi.org/10.1038/nphys3933} {\bibfield  {journal} {\bibinfo  {journal} {Nature Phys}\ }\textbf {\bibinfo {volume} {13}},\ \bibinfo {pages} {300} (\bibinfo {year} {2017}{\natexlab{b}})}\BibitemShut {NoStop}%
\bibitem [{\citenamefont {Shi}\ \emph {et~al.}(2019)\citenamefont {Shi}, \citenamefont {Liang}, \citenamefont {Zhu}, \citenamefont {Cai}, \citenamefont {Pollard}, \citenamefont {Wang}, \citenamefont {Wang}, \citenamefont {Wang}, \citenamefont {He}, \citenamefont {Yu}, \citenamefont {Eda}, \citenamefont {Liang},\ and\ \citenamefont {Yang}}]{shi_all-electric_2019}%
  \BibitemOpen
  \bibfield  {author} {\bibinfo {author} {\bibfnamefont {S.}~\bibnamefont {Shi}}, \bibinfo {author} {\bibfnamefont {S.}~\bibnamefont {Liang}}, \bibinfo {author} {\bibfnamefont {Z.}~\bibnamefont {Zhu}}, \bibinfo {author} {\bibfnamefont {K.}~\bibnamefont {Cai}}, \bibinfo {author} {\bibfnamefont {S.~D.}\ \bibnamefont {Pollard}}, \bibinfo {author} {\bibfnamefont {Y.}~\bibnamefont {Wang}}, \bibinfo {author} {\bibfnamefont {J.}~\bibnamefont {Wang}}, \bibinfo {author} {\bibfnamefont {Q.}~\bibnamefont {Wang}}, \bibinfo {author} {\bibfnamefont {P.}~\bibnamefont {He}}, \bibinfo {author} {\bibfnamefont {J.}~\bibnamefont {Yu}}, \bibinfo {author} {\bibfnamefont {G.}~\bibnamefont {Eda}}, \bibinfo {author} {\bibfnamefont {G.}~\bibnamefont {Liang}},\ and\ \bibinfo {author} {\bibfnamefont {H.}~\bibnamefont {Yang}},\ }\bibfield  {title} {\bibinfo {title} {All-electric magnetization switching and dzyaloshinskii–moriya interaction in wte2/ferromagnet heterostructures},\ }\href {https://doi.org/10.1038/s41565-019-0525-8}
  {\bibfield  {journal} {\bibinfo  {journal} {Nature Nanotechnology}\ }\textbf {\bibinfo {volume} {14}},\ \bibinfo {pages} {945–949} (\bibinfo {year} {2019})}\BibitemShut {NoStop}%
\bibitem [{\citenamefont {Liu}\ \emph {et~al.}(2021)\citenamefont {Liu}, \citenamefont {Zhou}, \citenamefont {Shu}, \citenamefont {Li}, \citenamefont {Zhao}, \citenamefont {Lin}, \citenamefont {Deng}, \citenamefont {Xie}, \citenamefont {Chen}, \citenamefont {Zhou}, \citenamefont {Guo}, \citenamefont {Wang}, \citenamefont {Yu}, \citenamefont {Shi}, \citenamefont {Yang}, \citenamefont {Pennycook}, \citenamefont {Manchon},\ and\ \citenamefont {Chen}}]{liu_symmetry-dependent_2021}%
  \BibitemOpen
  \bibfield  {author} {\bibinfo {author} {\bibfnamefont {L.}~\bibnamefont {Liu}}, \bibinfo {author} {\bibfnamefont {C.}~\bibnamefont {Zhou}}, \bibinfo {author} {\bibfnamefont {X.}~\bibnamefont {Shu}}, \bibinfo {author} {\bibfnamefont {C.}~\bibnamefont {Li}}, \bibinfo {author} {\bibfnamefont {T.}~\bibnamefont {Zhao}}, \bibinfo {author} {\bibfnamefont {W.}~\bibnamefont {Lin}}, \bibinfo {author} {\bibfnamefont {J.}~\bibnamefont {Deng}}, \bibinfo {author} {\bibfnamefont {Q.}~\bibnamefont {Xie}}, \bibinfo {author} {\bibfnamefont {S.}~\bibnamefont {Chen}}, \bibinfo {author} {\bibfnamefont {J.}~\bibnamefont {Zhou}}, \bibinfo {author} {\bibfnamefont {R.}~\bibnamefont {Guo}}, \bibinfo {author} {\bibfnamefont {H.}~\bibnamefont {Wang}}, \bibinfo {author} {\bibfnamefont {J.}~\bibnamefont {Yu}}, \bibinfo {author} {\bibfnamefont {S.}~\bibnamefont {Shi}}, \bibinfo {author} {\bibfnamefont {P.}~\bibnamefont {Yang}}, \bibinfo {author} {\bibfnamefont {S.}~\bibnamefont {Pennycook}}, \bibinfo {author} {\bibfnamefont
  {A.}~\bibnamefont {Manchon}},\ and\ \bibinfo {author} {\bibfnamefont {J.}~\bibnamefont {Chen}},\ }\bibfield  {title} {\bibinfo {title} {Symmetry-dependent field-free switching of perpendicular magnetization},\ }\href {https://doi.org/10.1038/s41565-020-00826-8} {\bibfield  {journal} {\bibinfo  {journal} {Nat. Nanotechnol.}\ }\textbf {\bibinfo {volume} {16}},\ \bibinfo {pages} {277} (\bibinfo {year} {2021})}\BibitemShut {NoStop}%
\bibitem [{\citenamefont {Ye}\ \emph {et~al.}(2022)\citenamefont {Ye}, \citenamefont {Zhu}, \citenamefont {Xu}, \citenamefont {Shang}, \citenamefont {Liu},\ and\ \citenamefont {Liao}}]{ye_orbit-transfer_2022}%
  \BibitemOpen
  \bibfield  {author} {\bibinfo {author} {\bibfnamefont {X.-G.}\ \bibnamefont {Ye}}, \bibinfo {author} {\bibfnamefont {P.-F.}\ \bibnamefont {Zhu}}, \bibinfo {author} {\bibfnamefont {W.-Z.}\ \bibnamefont {Xu}}, \bibinfo {author} {\bibfnamefont {N.}~\bibnamefont {Shang}}, \bibinfo {author} {\bibfnamefont {K.}~\bibnamefont {Liu}},\ and\ \bibinfo {author} {\bibfnamefont {Z.-M.}\ \bibnamefont {Liao}},\ }\bibfield  {title} {\bibinfo {title} {Orbit-transfer torque driven field-free switching of perpendicular magnetization},\ }\href {https://doi.org/10.1088/0256-307x/39/3/037303} {\bibfield  {journal} {\bibinfo  {journal} {Chinese Physics Letters}\ }\textbf {\bibinfo {volume} {39}},\ \bibinfo {pages} {037303} (\bibinfo {year} {2022})}\BibitemShut {NoStop}%
\bibitem [{\citenamefont {Canonico}\ \emph {et~al.}(2023)\citenamefont {Canonico}, \citenamefont {García},\ and\ \citenamefont {Roche}}]{canonico_spin-orbit_2023}%
  \BibitemOpen
  \bibfield  {author} {\bibinfo {author} {\bibfnamefont {L.~M.}\ \bibnamefont {Canonico}}, \bibinfo {author} {\bibfnamefont {J.~H.}\ \bibnamefont {García}},\ and\ \bibinfo {author} {\bibfnamefont {S.}~\bibnamefont {Roche}},\ }\href {https://doi.org/10.48550/ARXIV.2307.14673} {\bibinfo {title} {Spin-orbit torque emerging from orbital textures in centrosymmetric materials}} (\bibinfo {year} {2023})\BibitemShut {NoStop}%
\bibitem [{\citenamefont {Johansson}\ \emph {et~al.}(2021)\citenamefont {Johansson}, \citenamefont {Gobel}, \citenamefont {Henk}, \citenamefont {Bibes},\ and\ \citenamefont {Mertig}}]{johansson_spin_2021}%
  \BibitemOpen
  \bibfield  {author} {\bibinfo {author} {\bibfnamefont {A.}~\bibnamefont {Johansson}}, \bibinfo {author} {\bibfnamefont {B.}~\bibnamefont {Gobel}}, \bibinfo {author} {\bibfnamefont {J.}~\bibnamefont {Henk}}, \bibinfo {author} {\bibfnamefont {M.}~\bibnamefont {Bibes}},\ and\ \bibinfo {author} {\bibfnamefont {I.}~\bibnamefont {Mertig}},\ }\bibfield  {title} {\bibinfo {title} {Spin and orbital {Edelstein} effects in a two-dimensional electron gas: {Theory} and application to {SrTiO} 3 interfaces},\ }\href {https://doi.org/10.1103/PhysRevResearch.3.013275} {\bibfield  {journal} {\bibinfo  {journal} {Physical Review Research}\ }\textbf {\bibinfo {volume} {3}},\ \bibinfo {pages} {013275} (\bibinfo {year} {2021})}\BibitemShut {NoStop}%
\bibitem [{\citenamefont {Mecklenburg}\ and\ \citenamefont {Regan}(2011)}]{mecklenburg_spin_2011}%
  \BibitemOpen
  \bibfield  {author} {\bibinfo {author} {\bibfnamefont {M.}~\bibnamefont {Mecklenburg}}\ and\ \bibinfo {author} {\bibfnamefont {B.~C.}\ \bibnamefont {Regan}},\ }\bibfield  {title} {\bibinfo {title} {Spin and the honeycomb lattice: Lessons from graphene},\ }\href {https://doi.org/10.1103/physrevlett.106.116803} {\bibfield  {journal} {\bibinfo  {journal} {Physical Review Letters}\ }\textbf {\bibinfo {volume} {106}},\ \bibinfo {pages} {116803} (\bibinfo {year} {2011})}\BibitemShut {NoStop}%
\bibitem [{\citenamefont {Tuan}\ \emph {et~al.}(2014)\citenamefont {Tuan}, \citenamefont {Ortmann}, \citenamefont {Soriano}, \citenamefont {Valenzuela},\ and\ \citenamefont {Roche}}]{tuan_pseudospin-driven_2014}%
  \BibitemOpen
  \bibfield  {author} {\bibinfo {author} {\bibfnamefont {D.~V.}\ \bibnamefont {Tuan}}, \bibinfo {author} {\bibfnamefont {F.}~\bibnamefont {Ortmann}}, \bibinfo {author} {\bibfnamefont {D.}~\bibnamefont {Soriano}}, \bibinfo {author} {\bibfnamefont {S.~O.}\ \bibnamefont {Valenzuela}},\ and\ \bibinfo {author} {\bibfnamefont {S.}~\bibnamefont {Roche}},\ }\bibfield  {title} {\bibinfo {title} {Pseudospin-driven spin relaxation mechanism in graphene},\ }\href {https://doi.org/10.1038/nphys3083} {\bibfield  {journal} {\bibinfo  {journal} {Nature Phys.}\ }\textbf {\bibinfo {volume} {10}},\ \bibinfo {pages} {857} (\bibinfo {year} {2014})}\BibitemShut {NoStop}%
\bibitem [{\citenamefont {Offidani}\ \emph {et~al.}(2017)\citenamefont {Offidani}, \citenamefont {Milletar{\`i}}, \citenamefont {Raimondi},\ and\ \citenamefont {Ferreira}}]{offidani_optimal_2017}%
  \BibitemOpen
  \bibfield  {author} {\bibinfo {author} {\bibfnamefont {M.}~\bibnamefont {Offidani}}, \bibinfo {author} {\bibfnamefont {M.}~\bibnamefont {Milletar{\`i}}}, \bibinfo {author} {\bibfnamefont {R.}~\bibnamefont {Raimondi}},\ and\ \bibinfo {author} {\bibfnamefont {A.}~\bibnamefont {Ferreira}},\ }\bibfield  {title} {\bibinfo {title} {Optimal charge-to-spin conversion in graphene on transition-metal dichalcogenides},\ }\href {https://doi.org/10.1103/PhysRevLett.119.196801} {\bibfield  {journal} {\bibinfo  {journal} {Phys. Rev. Lett.}\ }\textbf {\bibinfo {volume} {119}},\ \bibinfo {pages} {196801} (\bibinfo {year} {2017})}\BibitemShut {NoStop}%
\bibitem [{\citenamefont {de~Moraes}\ \emph {et~al.}(2020)\citenamefont {de~Moraes}, \citenamefont {Cummings},\ and\ \citenamefont {Roche}}]{de-moraes_emergence_2020}%
  \BibitemOpen
  \bibfield  {author} {\bibinfo {author} {\bibfnamefont {B.~G.}\ \bibnamefont {de~Moraes}}, \bibinfo {author} {\bibfnamefont {A.~W.}\ \bibnamefont {Cummings}},\ and\ \bibinfo {author} {\bibfnamefont {S.}~\bibnamefont {Roche}},\ }\bibfield  {title} {\bibinfo {title} {Emergence of intraparticle entanglement and time-varying violation of {Bell}'s inequality in {Dirac} matter},\ }\href {https://doi.org/10.1103/PhysRevB.102.041403} {\bibfield  {journal} {\bibinfo  {journal} {Phys. Rev. B}\ }\textbf {\bibinfo {volume} {102}},\ \bibinfo {pages} {041403(R)} (\bibinfo {year} {2020})}\BibitemShut {NoStop}%
\bibitem [{sup()}]{supmat}%
  \BibitemOpen
  \href@noop {} {}\bibinfo {note} {See the Supplemental Material at [URL] for details on the transport methodology and detailed calculations of the band structure and transport properties of the systems.}\BibitemShut {Stop}%
\bibitem [{\citenamefont {Khokhriakov}\ \emph {et~al.}(2018)\citenamefont {Khokhriakov}, \citenamefont {Cummings}, \citenamefont {Song}, \citenamefont {Vila}, \citenamefont {Karpiak}, \citenamefont {Dankert}, \citenamefont {Roche},\ and\ \citenamefont {Dash}}]{khokhriakov_tailoring_2018}%
  \BibitemOpen
  \bibfield  {author} {\bibinfo {author} {\bibfnamefont {D.}~\bibnamefont {Khokhriakov}}, \bibinfo {author} {\bibfnamefont {A.~W.}\ \bibnamefont {Cummings}}, \bibinfo {author} {\bibfnamefont {K.}~\bibnamefont {Song}}, \bibinfo {author} {\bibfnamefont {M.}~\bibnamefont {Vila}}, \bibinfo {author} {\bibfnamefont {B.}~\bibnamefont {Karpiak}}, \bibinfo {author} {\bibfnamefont {A.}~\bibnamefont {Dankert}}, \bibinfo {author} {\bibfnamefont {S.}~\bibnamefont {Roche}},\ and\ \bibinfo {author} {\bibfnamefont {S.~P.}\ \bibnamefont {Dash}},\ }\bibfield  {title} {\bibinfo {title} {Tailoring emergent spin phenomena in dirac material heterostructures},\ }\bibfield  {journal} {\bibinfo  {journal} {Science Advances}\ }\textbf {\bibinfo {volume} {4}},\ \href {https://doi.org/10.1126/sciadv.aat9349} {10.1126/sciadv.aat9349} (\bibinfo {year} {2018})\BibitemShut {NoStop}%
\bibitem [{\citenamefont {Lee}\ \emph {et~al.}(2015)\citenamefont {Lee}, \citenamefont {Jin}, \citenamefont {Sung}, \citenamefont {Kim}, \citenamefont {Ryu}, \citenamefont {Park}, \citenamefont {Jhi}, \citenamefont {Kim}, \citenamefont {Kim}, \citenamefont {Yu}, \citenamefont {Kim}, \citenamefont {Noh},\ and\ \citenamefont {Chung}}]{lee_proximity_2015}%
  \BibitemOpen
  \bibfield  {author} {\bibinfo {author} {\bibfnamefont {P.}~\bibnamefont {Lee}}, \bibinfo {author} {\bibfnamefont {K.-H.}\ \bibnamefont {Jin}}, \bibinfo {author} {\bibfnamefont {S.~J.}\ \bibnamefont {Sung}}, \bibinfo {author} {\bibfnamefont {J.~G.}\ \bibnamefont {Kim}}, \bibinfo {author} {\bibfnamefont {M.-T.}\ \bibnamefont {Ryu}}, \bibinfo {author} {\bibfnamefont {H.-M.}\ \bibnamefont {Park}}, \bibinfo {author} {\bibfnamefont {S.-H.}\ \bibnamefont {Jhi}}, \bibinfo {author} {\bibfnamefont {N.}~\bibnamefont {Kim}}, \bibinfo {author} {\bibfnamefont {Y.}~\bibnamefont {Kim}}, \bibinfo {author} {\bibfnamefont {S.~U.}\ \bibnamefont {Yu}}, \bibinfo {author} {\bibfnamefont {K.~S.}\ \bibnamefont {Kim}}, \bibinfo {author} {\bibfnamefont {D.~Y.}\ \bibnamefont {Noh}},\ and\ \bibinfo {author} {\bibfnamefont {J.}~\bibnamefont {Chung}},\ }\bibfield  {title} {\bibinfo {title} {Proximity effect induced electronic properties of graphene on {Bi2}{Te2}{Se}},\ }\href {https://doi.org/10.1021/acsnano.5b03821} {\bibfield
  {journal} {\bibinfo  {journal} {ACS Nano}\ }\textbf {\bibinfo {volume} {9}},\ \bibinfo {pages} {10861–10866} (\bibinfo {year} {2015})}\BibitemShut {NoStop}%
\bibitem [{\citenamefont {Song}\ \emph {et~al.}(2018{\natexlab{a}})\citenamefont {Song}, \citenamefont {Soriano}, \citenamefont {Cummings}, \citenamefont {Robles}, \citenamefont {Ordejón},\ and\ \citenamefont {Roche}}]{song_spin_2018}%
  \BibitemOpen
  \bibfield  {author} {\bibinfo {author} {\bibfnamefont {K.}~\bibnamefont {Song}}, \bibinfo {author} {\bibfnamefont {D.}~\bibnamefont {Soriano}}, \bibinfo {author} {\bibfnamefont {A.~W.}\ \bibnamefont {Cummings}}, \bibinfo {author} {\bibfnamefont {R.}~\bibnamefont {Robles}}, \bibinfo {author} {\bibfnamefont {P.}~\bibnamefont {Ordejón}},\ and\ \bibinfo {author} {\bibfnamefont {S.}~\bibnamefont {Roche}},\ }\bibfield  {title} {\bibinfo {title} {Spin proximity effects in graphene/topological insulator heterostructures},\ }\href {https://doi.org/10.1021/acs.nanolett.7b05482} {\bibfield  {journal} {\bibinfo  {journal} {Nano Letters}\ }\textbf {\bibinfo {volume} {18}},\ \bibinfo {pages} {2033–2039} (\bibinfo {year} {2018}{\natexlab{a}})}\BibitemShut {NoStop}%
\bibitem [{\citenamefont {Naimer}\ and\ \citenamefont {Fabian}(2023)}]{naimer_twist-angle_2023}%
  \BibitemOpen
  \bibfield  {author} {\bibinfo {author} {\bibfnamefont {T.}~\bibnamefont {Naimer}}\ and\ \bibinfo {author} {\bibfnamefont {J.}~\bibnamefont {Fabian}},\ }\bibfield  {title} {\bibinfo {title} {Twist-angle dependent proximity induced spin-orbit coupling in graphene/topological insulator heterostructures},\ }\href {https://doi.org/10.1103/physrevb.107.195144} {\bibfield  {journal} {\bibinfo  {journal} {Physical Review B}\ }\textbf {\bibinfo {volume} {107}},\ \bibinfo {pages} {195144} (\bibinfo {year} {2023})}\BibitemShut {NoStop}%
\bibitem [{\citenamefont {Jin}\ and\ \citenamefont {Jhi}(2013)}]{jin_proximity-induced_2013}%
  \BibitemOpen
  \bibfield  {author} {\bibinfo {author} {\bibfnamefont {K.-H.}\ \bibnamefont {Jin}}\ and\ \bibinfo {author} {\bibfnamefont {S.-H.}\ \bibnamefont {Jhi}},\ }\bibfield  {title} {\bibinfo {title} {Proximity-induced giant spin-orbit interaction in epitaxial graphene on a topological insulator},\ }\href {https://doi.org/10.1103/PhysRevB.87.075442} {\bibfield  {journal} {\bibinfo  {journal} {Phys. Rev. B}\ }\textbf {\bibinfo {volume} {87}},\ \bibinfo {pages} {075442} (\bibinfo {year} {2013})}\BibitemShut {NoStop}%
\bibitem [{\citenamefont {Wakamura}\ \emph {et~al.}(2018)\citenamefont {Wakamura}, \citenamefont {Reale}, \citenamefont {Palczynski}, \citenamefont {Gu\'eron}, \citenamefont {Mattevi},\ and\ \citenamefont {Bouchiat}}]{wakamura_strong_2018}%
  \BibitemOpen
  \bibfield  {author} {\bibinfo {author} {\bibfnamefont {T.}~\bibnamefont {Wakamura}}, \bibinfo {author} {\bibfnamefont {F.}~\bibnamefont {Reale}}, \bibinfo {author} {\bibfnamefont {P.}~\bibnamefont {Palczynski}}, \bibinfo {author} {\bibfnamefont {S.}~\bibnamefont {Gu\'eron}}, \bibinfo {author} {\bibfnamefont {C.}~\bibnamefont {Mattevi}},\ and\ \bibinfo {author} {\bibfnamefont {H.}~\bibnamefont {Bouchiat}},\ }\bibfield  {title} {\bibinfo {title} {Strong anisotropic spin-orbit interaction induced in graphene by monolayer ${\mathrm{ws}}_{2}$},\ }\href {https://doi.org/10.1103/PhysRevLett.120.106802} {\bibfield  {journal} {\bibinfo  {journal} {Phys. Rev. Lett.}\ }\textbf {\bibinfo {volume} {120}},\ \bibinfo {pages} {106802} (\bibinfo {year} {2018})}\BibitemShut {NoStop}%
\bibitem [{\citenamefont {Wakamura}\ \emph {et~al.}(2019)\citenamefont {Wakamura}, \citenamefont {Reale}, \citenamefont {Palczynski}, \citenamefont {Zhao}, \citenamefont {Johnson}, \citenamefont {Gu\'eron}, \citenamefont {Mattevi}, \citenamefont {Ouerghi},\ and\ \citenamefont {Bouchiat}}]{wakamura_spin-orbit_2019}%
  \BibitemOpen
  \bibfield  {author} {\bibinfo {author} {\bibfnamefont {T.}~\bibnamefont {Wakamura}}, \bibinfo {author} {\bibfnamefont {F.}~\bibnamefont {Reale}}, \bibinfo {author} {\bibfnamefont {P.}~\bibnamefont {Palczynski}}, \bibinfo {author} {\bibfnamefont {M.~Q.}\ \bibnamefont {Zhao}}, \bibinfo {author} {\bibfnamefont {A.~T.~C.}\ \bibnamefont {Johnson}}, \bibinfo {author} {\bibfnamefont {S.}~\bibnamefont {Gu\'eron}}, \bibinfo {author} {\bibfnamefont {C.}~\bibnamefont {Mattevi}}, \bibinfo {author} {\bibfnamefont {A.}~\bibnamefont {Ouerghi}},\ and\ \bibinfo {author} {\bibfnamefont {H.}~\bibnamefont {Bouchiat}},\ }\bibfield  {title} {\bibinfo {title} {Spin-orbit interaction induced in graphene by transition metal dichalcogenides},\ }\href {https://doi.org/10.1103/PhysRevB.99.245402} {\bibfield  {journal} {\bibinfo  {journal} {Phys. Rev. B}\ }\textbf {\bibinfo {volume} {99}},\ \bibinfo {pages} {245402} (\bibinfo {year} {2019})}\BibitemShut {NoStop}%
\bibitem [{\citenamefont {Avsar}\ \emph {et~al.}(2014)\citenamefont {Avsar}, \citenamefont {Tan}, \citenamefont {Taychatanapat}, \citenamefont {Balakrishnan}, \citenamefont {Koon}, \citenamefont {Yeo}, \citenamefont {Lahiri}, \citenamefont {Carvalho}, \citenamefont {Rodin}, \citenamefont {O’Farrell}, \citenamefont {Eda}, \citenamefont {Castro~Neto},\ and\ \citenamefont {{\"O}zyilmaz}}]{avsar_spin-orbit_2014}%
  \BibitemOpen
  \bibfield  {author} {\bibinfo {author} {\bibfnamefont {A.}~\bibnamefont {Avsar}}, \bibinfo {author} {\bibfnamefont {J.~Y.}\ \bibnamefont {Tan}}, \bibinfo {author} {\bibfnamefont {T.}~\bibnamefont {Taychatanapat}}, \bibinfo {author} {\bibfnamefont {J.}~\bibnamefont {Balakrishnan}}, \bibinfo {author} {\bibfnamefont {G.~K.~W.}\ \bibnamefont {Koon}}, \bibinfo {author} {\bibfnamefont {Y.}~\bibnamefont {Yeo}}, \bibinfo {author} {\bibfnamefont {J.}~\bibnamefont {Lahiri}}, \bibinfo {author} {\bibfnamefont {A.}~\bibnamefont {Carvalho}}, \bibinfo {author} {\bibfnamefont {A.~S.}\ \bibnamefont {Rodin}}, \bibinfo {author} {\bibfnamefont {E.~C.~T.}\ \bibnamefont {O’Farrell}}, \bibinfo {author} {\bibfnamefont {G.}~\bibnamefont {Eda}}, \bibinfo {author} {\bibfnamefont {A.~H.}\ \bibnamefont {Castro~Neto}},\ and\ \bibinfo {author} {\bibfnamefont {B.}~\bibnamefont {{\"O}zyilmaz}},\ }\bibfield  {title} {\bibinfo {title} {Spin–orbit proximity effect in graphene},\ }\href {https://doi.org/10.1038/ncomms5875} {\bibfield
  {journal} {\bibinfo  {journal} {Nature Commun.}\ }\textbf {\bibinfo {volume} {5}},\ \bibinfo {pages} {4875} (\bibinfo {year} {2014})}\BibitemShut {NoStop}%
\bibitem [{\citenamefont {Wang}\ \emph {et~al.}(2015)\citenamefont {Wang}, \citenamefont {Ki}, \citenamefont {Chen}, \citenamefont {Berger}, \citenamefont {MacDonald},\ and\ \citenamefont {Morpurgo}}]{wang_strong_2015}%
  \BibitemOpen
  \bibfield  {author} {\bibinfo {author} {\bibfnamefont {Z.}~\bibnamefont {Wang}}, \bibinfo {author} {\bibfnamefont {D.-K.}\ \bibnamefont {Ki}}, \bibinfo {author} {\bibfnamefont {H.}~\bibnamefont {Chen}}, \bibinfo {author} {\bibfnamefont {H.}~\bibnamefont {Berger}}, \bibinfo {author} {\bibfnamefont {A.~H.}\ \bibnamefont {MacDonald}},\ and\ \bibinfo {author} {\bibfnamefont {A.~F.}\ \bibnamefont {Morpurgo}},\ }\bibfield  {title} {\bibinfo {title} {Strong interface-induced spin–orbit interaction in graphene on {${\mathrm{WS}}_{2}$}},\ }\href {https://doi.org/10.1038/ncomms9339} {\bibfield  {journal} {\bibinfo  {journal} {Nature Commun.}\ }\textbf {\bibinfo {volume} {6}},\ \bibinfo {pages} {8339} (\bibinfo {year} {2015})}\BibitemShut {NoStop}%
\bibitem [{\citenamefont {Wang}\ \emph {et~al.}(2016)\citenamefont {Wang}, \citenamefont {Ki}, \citenamefont {Khoo}, \citenamefont {Mauro}, \citenamefont {Berger}, \citenamefont {Levitov},\ and\ \citenamefont {Morpurgo}}]{wang_origin_2016}%
  \BibitemOpen
  \bibfield  {author} {\bibinfo {author} {\bibfnamefont {Z.}~\bibnamefont {Wang}}, \bibinfo {author} {\bibfnamefont {D.-K.}\ \bibnamefont {Ki}}, \bibinfo {author} {\bibfnamefont {J.~Y.}\ \bibnamefont {Khoo}}, \bibinfo {author} {\bibfnamefont {D.}~\bibnamefont {Mauro}}, \bibinfo {author} {\bibfnamefont {H.}~\bibnamefont {Berger}}, \bibinfo {author} {\bibfnamefont {L.~S.}\ \bibnamefont {Levitov}},\ and\ \bibinfo {author} {\bibfnamefont {A.~F.}\ \bibnamefont {Morpurgo}},\ }\bibfield  {title} {\bibinfo {title} {Origin and magnitude of `designer' spin-orbit interaction in graphene on semiconducting transition metal dichalcogenides},\ }\href {https://doi.org/10.1103/PhysRevX.6.041020} {\bibfield  {journal} {\bibinfo  {journal} {Phys. Rev. X}\ }\textbf {\bibinfo {volume} {6}},\ \bibinfo {pages} {041020} (\bibinfo {year} {2016})}\BibitemShut {NoStop}%
\bibitem [{\citenamefont {Gmitra}\ and\ \citenamefont {Fabian}(2015)}]{gmitra_graphene_2015}%
  \BibitemOpen
  \bibfield  {author} {\bibinfo {author} {\bibfnamefont {M.}~\bibnamefont {Gmitra}}\ and\ \bibinfo {author} {\bibfnamefont {J.}~\bibnamefont {Fabian}},\ }\bibfield  {title} {\bibinfo {title} {Graphene on transition-metal dichalcogenides: {A} platform for proximity spin-orbit physics and optospintronics},\ }\href {https://doi.org/10.1103/PhysRevB.92.155403} {\bibfield  {journal} {\bibinfo  {journal} {Phys. Rev. B}\ }\textbf {\bibinfo {volume} {92}},\ \bibinfo {pages} {155403} (\bibinfo {year} {2015})}\BibitemShut {NoStop}%
\bibitem [{\citenamefont {Tang}\ \emph {et~al.}(2017)\citenamefont {Tang}, \citenamefont {Cheng}, \citenamefont {Aldosary}, \citenamefont {Wang}, \citenamefont {Jiang}, \citenamefont {Watanabe}, \citenamefont {Taniguchi}, \citenamefont {Bockrath},\ and\ \citenamefont {Shi}}]{tang_approaching_2017}%
  \BibitemOpen
  \bibfield  {author} {\bibinfo {author} {\bibfnamefont {C.}~\bibnamefont {Tang}}, \bibinfo {author} {\bibfnamefont {B.}~\bibnamefont {Cheng}}, \bibinfo {author} {\bibfnamefont {M.}~\bibnamefont {Aldosary}}, \bibinfo {author} {\bibfnamefont {Z.}~\bibnamefont {Wang}}, \bibinfo {author} {\bibfnamefont {Z.}~\bibnamefont {Jiang}}, \bibinfo {author} {\bibfnamefont {K.}~\bibnamefont {Watanabe}}, \bibinfo {author} {\bibfnamefont {T.}~\bibnamefont {Taniguchi}}, \bibinfo {author} {\bibfnamefont {M.}~\bibnamefont {Bockrath}},\ and\ \bibinfo {author} {\bibfnamefont {J.}~\bibnamefont {Shi}},\ }\bibfield  {title} {\bibinfo {title} {Approaching quantum anomalous {Hall} effect in proximity-coupled {YIG}/{graphene}/{h-BN} sandwich structure},\ }\bibfield  {journal} {\bibinfo  {journal} {APL Materials}\ }\textbf {\bibinfo {volume} {6}},\ \href {https://doi.org/10.1063/1.5001318} {10.1063/1.5001318} (\bibinfo {year} {2017})\BibitemShut {NoStop}%
\bibitem [{\citenamefont {Wei}\ \emph {et~al.}(2016)\citenamefont {Wei}, \citenamefont {Lee}, \citenamefont {Lemaitre}, \citenamefont {Pinel}, \citenamefont {Cutaia}, \citenamefont {Cha}, \citenamefont {Katmis}, \citenamefont {Zhu}, \citenamefont {Heiman}, \citenamefont {Hone}, \citenamefont {Moodera},\ and\ \citenamefont {Chen}}]{wei_strong_2016}%
  \BibitemOpen
  \bibfield  {author} {\bibinfo {author} {\bibfnamefont {P.}~\bibnamefont {Wei}}, \bibinfo {author} {\bibfnamefont {S.}~\bibnamefont {Lee}}, \bibinfo {author} {\bibfnamefont {F.}~\bibnamefont {Lemaitre}}, \bibinfo {author} {\bibfnamefont {L.}~\bibnamefont {Pinel}}, \bibinfo {author} {\bibfnamefont {D.}~\bibnamefont {Cutaia}}, \bibinfo {author} {\bibfnamefont {W.}~\bibnamefont {Cha}}, \bibinfo {author} {\bibfnamefont {F.}~\bibnamefont {Katmis}}, \bibinfo {author} {\bibfnamefont {Y.}~\bibnamefont {Zhu}}, \bibinfo {author} {\bibfnamefont {D.}~\bibnamefont {Heiman}}, \bibinfo {author} {\bibfnamefont {J.}~\bibnamefont {Hone}}, \bibinfo {author} {\bibfnamefont {J.~S.}\ \bibnamefont {Moodera}},\ and\ \bibinfo {author} {\bibfnamefont {C.-T.}\ \bibnamefont {Chen}},\ }\bibfield  {title} {\bibinfo {title} {Strong interfacial exchange field in the {graphene}/{EuS} heterostructure},\ }\href {https://doi.org/10.1038/nmat4603} {\bibfield  {journal} {\bibinfo  {journal} {Nature Materials}\ }\textbf {\bibinfo {volume} {15}},\
  \bibinfo {pages} {711–716} (\bibinfo {year} {2016})}\BibitemShut {NoStop}%
\bibitem [{\citenamefont {Wu}\ \emph {et~al.}(2017)\citenamefont {Wu}, \citenamefont {Song}, \citenamefont {Zhang}, \citenamefont {Yang}, \citenamefont {Ren}, \citenamefont {Liu}, \citenamefont {Wu}, \citenamefont {Wu}, \citenamefont {Li}, \citenamefont {Jia}, \citenamefont {Yan}, \citenamefont {Wu}, \citenamefont {Duan}, \citenamefont {Han}, \citenamefont {Liao},\ and\ \citenamefont {Yu}}]{wu_magnetic_2017}%
  \BibitemOpen
  \bibfield  {author} {\bibinfo {author} {\bibfnamefont {Y.-F.}\ \bibnamefont {Wu}}, \bibinfo {author} {\bibfnamefont {H.-D.}\ \bibnamefont {Song}}, \bibinfo {author} {\bibfnamefont {L.}~\bibnamefont {Zhang}}, \bibinfo {author} {\bibfnamefont {X.}~\bibnamefont {Yang}}, \bibinfo {author} {\bibfnamefont {Z.}~\bibnamefont {Ren}}, \bibinfo {author} {\bibfnamefont {D.}~\bibnamefont {Liu}}, \bibinfo {author} {\bibfnamefont {H.-C.}\ \bibnamefont {Wu}}, \bibinfo {author} {\bibfnamefont {J.}~\bibnamefont {Wu}}, \bibinfo {author} {\bibfnamefont {J.-G.}\ \bibnamefont {Li}}, \bibinfo {author} {\bibfnamefont {Z.}~\bibnamefont {Jia}}, \bibinfo {author} {\bibfnamefont {B.}~\bibnamefont {Yan}}, \bibinfo {author} {\bibfnamefont {X.}~\bibnamefont {Wu}}, \bibinfo {author} {\bibfnamefont {C.-G.}\ \bibnamefont {Duan}}, \bibinfo {author} {\bibfnamefont {G.}~\bibnamefont {Han}}, \bibinfo {author} {\bibfnamefont {Z.-M.}\ \bibnamefont {Liao}},\ and\ \bibinfo {author} {\bibfnamefont {D.}~\bibnamefont {Yu}},\ }\bibfield  {title}
  {\bibinfo {title} {Magnetic proximity effect in graphene coupled to a $\mathrm{BiFe}{\mathrm{o}}_{3}$ nanoplate},\ }\href {https://doi.org/10.1103/PhysRevB.95.195426} {\bibfield  {journal} {\bibinfo  {journal} {Phys. Rev. B}\ }\textbf {\bibinfo {volume} {95}},\ \bibinfo {pages} {195426} (\bibinfo {year} {2017})}\BibitemShut {NoStop}%
\bibitem [{\citenamefont {Song}\ \emph {et~al.}(2018{\natexlab{b}})\citenamefont {Song}, \citenamefont {Wu}, \citenamefont {Yang}, \citenamefont {Ren}, \citenamefont {Ke}, \citenamefont {Kurttepeli}, \citenamefont {Tendeloo}, \citenamefont {Liu}, \citenamefont {Wu}, \citenamefont {Yan}, \citenamefont {Wu}, \citenamefont {Duan}, \citenamefont {Han}, \citenamefont {Liao},\ and\ \citenamefont {Yu}}]{song_assymetric_2018}%
  \BibitemOpen
  \bibfield  {author} {\bibinfo {author} {\bibfnamefont {H.-D.}\ \bibnamefont {Song}}, \bibinfo {author} {\bibfnamefont {Y.-F.}\ \bibnamefont {Wu}}, \bibinfo {author} {\bibfnamefont {X.}~\bibnamefont {Yang}}, \bibinfo {author} {\bibfnamefont {Z.}~\bibnamefont {Ren}}, \bibinfo {author} {\bibfnamefont {X.}~\bibnamefont {Ke}}, \bibinfo {author} {\bibfnamefont {M.}~\bibnamefont {Kurttepeli}}, \bibinfo {author} {\bibfnamefont {G.~V.}\ \bibnamefont {Tendeloo}}, \bibinfo {author} {\bibfnamefont {D.}~\bibnamefont {Liu}}, \bibinfo {author} {\bibfnamefont {H.-C.}\ \bibnamefont {Wu}}, \bibinfo {author} {\bibfnamefont {B.}~\bibnamefont {Yan}}, \bibinfo {author} {\bibfnamefont {X.}~\bibnamefont {Wu}}, \bibinfo {author} {\bibfnamefont {C.-G.}\ \bibnamefont {Duan}}, \bibinfo {author} {\bibfnamefont {G.}~\bibnamefont {Han}}, \bibinfo {author} {\bibfnamefont {Z.-M.}\ \bibnamefont {Liao}},\ and\ \bibinfo {author} {\bibfnamefont {D.}~\bibnamefont {Yu}},\ }\bibfield  {title} {\bibinfo {title} {Asymmetric modulation on exchange
  field in a graphene/{BiFeO3} heterostructure by external magnetic field},\ }\href {https://doi.org/10.1021/acs.nanolett.7b05480} {\bibfield  {journal} {\bibinfo  {journal} {Nano Letters}\ }\textbf {\bibinfo {volume} {18}},\ \bibinfo {pages} {2435–2441} (\bibinfo {year} {2018}{\natexlab{b}})}\BibitemShut {NoStop}%
\bibitem [{\citenamefont {Wu}\ \emph {et~al.}(2020)\citenamefont {Wu}, \citenamefont {Yin}, \citenamefont {Pan}, \citenamefont {Grutter}, \citenamefont {Pan}, \citenamefont {Lee}, \citenamefont {Gilbert}, \citenamefont {Borchers}, \citenamefont {Ratcliff}, \citenamefont {Li}, \citenamefont {Han},\ and\ \citenamefont {Wang}}]{wu_large_2020}%
  \BibitemOpen
  \bibfield  {author} {\bibinfo {author} {\bibfnamefont {Y.}~\bibnamefont {Wu}}, \bibinfo {author} {\bibfnamefont {G.}~\bibnamefont {Yin}}, \bibinfo {author} {\bibfnamefont {L.}~\bibnamefont {Pan}}, \bibinfo {author} {\bibfnamefont {A.~J.}\ \bibnamefont {Grutter}}, \bibinfo {author} {\bibfnamefont {Q.}~\bibnamefont {Pan}}, \bibinfo {author} {\bibfnamefont {A.}~\bibnamefont {Lee}}, \bibinfo {author} {\bibfnamefont {D.~A.}\ \bibnamefont {Gilbert}}, \bibinfo {author} {\bibfnamefont {J.~A.}\ \bibnamefont {Borchers}}, \bibinfo {author} {\bibfnamefont {W.}~\bibnamefont {Ratcliff}}, \bibinfo {author} {\bibfnamefont {A.}~\bibnamefont {Li}}, \bibinfo {author} {\bibfnamefont {X.-d.}\ \bibnamefont {Han}},\ and\ \bibinfo {author} {\bibfnamefont {K.~L.}\ \bibnamefont {Wang}},\ }\bibfield  {title} {\bibinfo {title} {Large exchange splitting in monolayer graphene magnetized by an antiferromagnet},\ }\href {https://doi.org/10.1038/s41928-020-0458-0} {\bibfield  {journal} {\bibinfo  {journal} {Nature Electronics}\ }\textbf
  {\bibinfo {volume} {3}},\ \bibinfo {pages} {604–611} (\bibinfo {year} {2020})}\BibitemShut {NoStop}%
\bibitem [{\citenamefont {Ibrahim}\ \emph {et~al.}(2019)\citenamefont {Ibrahim}, \citenamefont {Hallal}, \citenamefont {Lerma}, \citenamefont {Waintal}, \citenamefont {Tsymbal},\ and\ \citenamefont {Chshiev}}]{ibrahim_unveiling_2019}%
  \BibitemOpen
  \bibfield  {author} {\bibinfo {author} {\bibfnamefont {F.}~\bibnamefont {Ibrahim}}, \bibinfo {author} {\bibfnamefont {A.}~\bibnamefont {Hallal}}, \bibinfo {author} {\bibfnamefont {D.~S.}\ \bibnamefont {Lerma}}, \bibinfo {author} {\bibfnamefont {X.}~\bibnamefont {Waintal}}, \bibinfo {author} {\bibfnamefont {E.~Y.}\ \bibnamefont {Tsymbal}},\ and\ \bibinfo {author} {\bibfnamefont {M.}~\bibnamefont {Chshiev}},\ }\bibfield  {title} {\bibinfo {title} {Unveiling multiferroic proximity effect in graphene},\ }\href {https://doi.org/10.1088/2053-1583/ab5319} {\bibfield  {journal} {\bibinfo  {journal} {2D Mater.}\ }\textbf {\bibinfo {volume} {7}},\ \bibinfo {pages} {015020} (\bibinfo {year} {2019})}\BibitemShut {NoStop}%
\bibitem [{\citenamefont {Qiao}\ \emph {et~al.}(2010)\citenamefont {Qiao}, \citenamefont {Yang}, \citenamefont {Feng}, \citenamefont {Tse}, \citenamefont {Ding}, \citenamefont {Yao}, \citenamefont {Wang},\ and\ \citenamefont {Niu}}]{qiao_quantum_2010}%
  \BibitemOpen
  \bibfield  {author} {\bibinfo {author} {\bibfnamefont {Z.}~\bibnamefont {Qiao}}, \bibinfo {author} {\bibfnamefont {S.~A.}\ \bibnamefont {Yang}}, \bibinfo {author} {\bibfnamefont {W.}~\bibnamefont {Feng}}, \bibinfo {author} {\bibfnamefont {W.-K.}\ \bibnamefont {Tse}}, \bibinfo {author} {\bibfnamefont {J.}~\bibnamefont {Ding}}, \bibinfo {author} {\bibfnamefont {Y.}~\bibnamefont {Yao}}, \bibinfo {author} {\bibfnamefont {J.}~\bibnamefont {Wang}},\ and\ \bibinfo {author} {\bibfnamefont {Q.}~\bibnamefont {Niu}},\ }\bibfield  {title} {\bibinfo {title} {Quantum anomalous {Hall} effect in graphene from {Rashba} and exchange effects},\ }\href {https://doi.org/10.1103/PhysRevB.82.161414} {\bibfield  {journal} {\bibinfo  {journal} {Phys. Rev. B}\ }\textbf {\bibinfo {volume} {82}},\ \bibinfo {pages} {161414(R)} (\bibinfo {year} {2010})}\BibitemShut {NoStop}%
\bibitem [{\citenamefont {Qiao}\ \emph {et~al.}(2014)\citenamefont {Qiao}, \citenamefont {Ren}, \citenamefont {Chen}, \citenamefont {Bellaiche}, \citenamefont {Zhang}, \citenamefont {MacDonald},\ and\ \citenamefont {Niu}}]{qiao_quantum_2014}%
  \BibitemOpen
  \bibfield  {author} {\bibinfo {author} {\bibfnamefont {Z.}~\bibnamefont {Qiao}}, \bibinfo {author} {\bibfnamefont {W.}~\bibnamefont {Ren}}, \bibinfo {author} {\bibfnamefont {H.}~\bibnamefont {Chen}}, \bibinfo {author} {\bibfnamefont {L.}~\bibnamefont {Bellaiche}}, \bibinfo {author} {\bibfnamefont {Z.}~\bibnamefont {Zhang}}, \bibinfo {author} {\bibfnamefont {A.~H.}\ \bibnamefont {MacDonald}},\ and\ \bibinfo {author} {\bibfnamefont {Q.}~\bibnamefont {Niu}},\ }\bibfield  {title} {\bibinfo {title} {Quantum anomalous {Hall} effect in graphene proximity coupled to an antiferromagnetic insulator},\ }\href {https://doi.org/10.1103/PhysRevLett.112.116404} {\bibfield  {journal} {\bibinfo  {journal} {Phys. Rev. Lett.}\ }\textbf {\bibinfo {volume} {112}},\ \bibinfo {pages} {116404} (\bibinfo {year} {2014})}\BibitemShut {NoStop}%
\bibitem [{\citenamefont {Xiao}\ \emph {et~al.}(2010)\citenamefont {Xiao}, \citenamefont {Chang},\ and\ \citenamefont {Niu}}]{xiao_berry_2010}%
  \BibitemOpen
  \bibfield  {author} {\bibinfo {author} {\bibfnamefont {D.}~\bibnamefont {Xiao}}, \bibinfo {author} {\bibfnamefont {M.-C.}\ \bibnamefont {Chang}},\ and\ \bibinfo {author} {\bibfnamefont {Q.}~\bibnamefont {Niu}},\ }\bibfield  {title} {\bibinfo {title} {Berry phase effects on electronic properties},\ }\href {https://doi.org/10.1103/RevModPhys.82.1959} {\bibfield  {journal} {\bibinfo  {journal} {Rev. Mod. Phys.}\ }\textbf {\bibinfo {volume} {82}},\ \bibinfo {pages} {1959} (\bibinfo {year} {2010})}\BibitemShut {NoStop}%
\bibitem [{\citenamefont {Go}\ \emph {et~al.}(2020)\citenamefont {Go}, \citenamefont {Freimuth}, \citenamefont {Hanke}, \citenamefont {Xue}, \citenamefont {Gomonay}, \citenamefont {Lee}, \citenamefont {Bl\"ugel}, \citenamefont {Haney}, \citenamefont {Lee},\ and\ \citenamefont {Mokrousov}}]{go_theory_2020}%
  \BibitemOpen
  \bibfield  {author} {\bibinfo {author} {\bibfnamefont {D.}~\bibnamefont {Go}}, \bibinfo {author} {\bibfnamefont {F.}~\bibnamefont {Freimuth}}, \bibinfo {author} {\bibfnamefont {J.-P.}\ \bibnamefont {Hanke}}, \bibinfo {author} {\bibfnamefont {F.}~\bibnamefont {Xue}}, \bibinfo {author} {\bibfnamefont {O.}~\bibnamefont {Gomonay}}, \bibinfo {author} {\bibfnamefont {K.-J.}\ \bibnamefont {Lee}}, \bibinfo {author} {\bibfnamefont {S.}~\bibnamefont {Bl\"ugel}}, \bibinfo {author} {\bibfnamefont {P.~M.}\ \bibnamefont {Haney}}, \bibinfo {author} {\bibfnamefont {H.-W.}\ \bibnamefont {Lee}},\ and\ \bibinfo {author} {\bibfnamefont {Y.}~\bibnamefont {Mokrousov}},\ }\bibfield  {title} {\bibinfo {title} {Theory of current-induced angular momentum transfer dynamics in spin-orbit coupled systems},\ }\href {https://doi.org/10.1103/PhysRevResearch.2.033401} {\bibfield  {journal} {\bibinfo  {journal} {Phys. Rev. Res.}\ }\textbf {\bibinfo {volume} {2}},\ \bibinfo {pages} {033401} (\bibinfo {year} {2020})}\BibitemShut {NoStop}%
\end{thebibliography}
\end{document}


\title{Supplemental Material for ``\JMD{Topologically Driven} Spin-Orbit Torque in Dirac Matter''}

\author{Joaqu{\'i}n Medina Due{\~n}as}
\affiliation{ICN2 --- Catalan Institute of Nanoscience and Nanotechnology, CSIC and BIST, Campus UAB, Bellaterra, 08193 Barcelona, Spain}
\affiliation{Department of Physics, Universitat Aut{\`o}noma de Barcelona (UAB), Campus UAB, Bellaterra, 08193 Barcelona, Spain}

\author{Jos{\'e} H. Garc{\'i}a}
\email{josehugo.garcia@icn2.cat}
\affiliation{ICN2 --- Catalan Institute of Nanoscience and Nanotechnology, CSIC and BIST, Campus UAB, Bellaterra, 08193 Barcelona, Spain}

\author{Stephan Roche}
\affiliation{ICN2 --- Catalan Institute of Nanoscience and Nanotechnology, CSIC and BIST,
Campus UAB, Bellaterra, 08193 Barcelona, Spain}
\affiliation{ICREA --- Instituci\'o Catalana de Recerca i Estudis Avan\c{c}ats, 08010 Barcelona, Spain}

\maketitle

\onecolumngrid

\section{Transport methodology}
We employ a semi-classical transport methodology based on Boltzmann transport theory, allowing us to understand the subjacent microscopic mechanisms generating the non-equilibrium spin density. Following the work introduced in Ref.~\cite{medina-duenas_emerging_2024}, the non-equilibrium spin density at Fermi level $\varepsilon_\text{F}$ is 
\begin{equation}
    \boldsymbol{S} (\varepsilon_\text{F}) = \sum_n \int \frac{\text{d}^2\boldsymbol{k}}{(2\pi)^2} \, \left[ \delta \mathit{f}_{n,\boldsymbol{k}} \boldsymbol{s}_{n,\boldsymbol{k}} + \mathit{f}(E_{n,\boldsymbol{k}}) \delta \boldsymbol{s}_{n,\boldsymbol{k}} \right] \text{ ,}
    \label{eq:noneqS}
\end{equation}
with $E_{n,\boldsymbol{k}}$ and $|E_{n,\boldsymbol{k}}\rangle$ the eigenvalue and eigenstate respectively of an electron with momentum $\boldsymbol{k}$ and band index $n$, and $\boldsymbol{s}_{n,\boldsymbol{k}} = \langle E_{n,\boldsymbol{k}} | \boldsymbol{s} | E_{n,\boldsymbol{k}} \rangle$ the spin texture. The first term in Eq.~\eqref{eq:noneqS} represents the standard Boltzmann transport contribution stemming from the current-induced variation of the carrier occupation, $\delta \mathit{f}_{n,\boldsymbol{k}}$, originated at the Fermi surface. 
\JMD{Under the momentum relaxation time approximation we obtain $\delta \mathit{f}_{n, \boldsymbol{k}} = - \delta(E_{n,\boldsymbol{k}} - \varepsilon_\text{F}) e\, \hbar^{-1}\, \tau\,  \boldsymbol{\mathcal{E}} \cdot \nabla_{\boldsymbol{k}} E_{n,\boldsymbol{k}}$, with $\tau$ the momentum relaxation time and $\boldsymbol{\mathcal{E}}$ the applied electric field, revealing that Fermi-surface transport properties diverge in the absence of disorder and require a stabilization mechanism impeding an infinite deformation of the carrier distribution.}
\JMD{On the other hand,} the second term \JMD{in Eq.~\eqref{eq:noneqS}} originates from the $k$-space transport of Bloch states under an electric field, which modifies the spin texture throughout the Fermi sea acquiring a non-equilibrium component $\delta \boldsymbol{s}_{n,\boldsymbol{k}}$, with $\mathit{f}_{n,\boldsymbol{k}}$ the equilibrium carrier occupation. While $\delta\boldsymbol{s}_{n,\boldsymbol{k}}$ may be computed from standard time-dependent perturbation theory \cite{xiao_berry_2010}, developing a semi-classical spin dynamics approach proved to provide a deeper grasp of the physics behind this contribution \cite{medina-duenas_emerging_2024}. The latter approach however, was only accurately developed in absence of spin-pseudospin entanglement. \JMD{We now develop a general method for computing the non-equilibrium spin texture from the coupled spin-pseudospin semi-classical dynamics.}

We consider a general hamiltonian of the form
\begin{equation}
    \mathcal{H}_{\boldsymbol{k}} = \mathcal{H}_{0,\boldsymbol{k}} - \frac{1}{2} B_i s_i - \frac{1}{2} \beta_j \sigma_j - \frac{1}{2} \lambda_{ij} s_i \sigma_j \text{ ,}
\end{equation}
with $\boldsymbol{B}$ and $\boldsymbol{\beta}$ the effective magnetic and pseudomagnetic fields respectively coupled to $\boldsymbol{s}$ and $\boldsymbol{\sigma}$, $\lambda_{ij}$ the spin-pseudospin coupling field, and $H_{0,\boldsymbol{k}}$ a scalar term which is spin and pseudospin independent, and where repeated indexes are implicitly summed over. The dynamics involve a set of 15 variables, $\langle s_i \rangle$, $\langle \sigma_j \rangle$ and $\langle \rho_{ij} \rangle = \langle s_i \sigma_j \rangle$, with $i,j=x,y,z$, whose dynamics are determined by the Ehrenfest theorem, yielding
\begin{subequations}
\begin{equation}
    \frac{\text{d}}{\text{d}t} \langle s_i \rangle = -\epsilon_{ijk} B_j \langle s_k \rangle - \epsilon_{ijk} \lambda_{jl} \langle \rho_{kl} \rangle \text{ ,}
\end{equation}
\begin{equation}
    \frac{\text{d}}{\text{d}t} \langle \sigma_i \rangle = -\epsilon_{ijk} \beta_j \langle s_k \rangle - \epsilon_{ijk} \lambda_{lj} \langle \rho_{lk} \rangle \text{ ,}
\end{equation}
\begin{equation}
    \frac{\text{d}}{\text{d}t} \langle \rho_{ij} \rangle = -\epsilon_{ikl} \lambda_{kj} \langle s_l \rangle - \epsilon_{jkl} \lambda_{ik} \langle \sigma_l \rangle - \epsilon_{ikl} B_k \langle \rho_{lj} \rangle - \epsilon_{jkl} \beta_k \langle \rho_{il} \rangle \text{ ,}
\end{equation}
\end{subequations}
where the band index and $\boldsymbol{k}$ dependence of the spin-pseudospin texture is implied.
We rephrase the problem as a matrix equation, for which we define the spin-pseudospin vector $\vec{x} = [ \dots \langle s_i \rangle \dots \langle \sigma_i \rangle \dots \langle \rho_{ij} \rangle \dots]$, and the Ehrenfest matrix $E$ such that $\frac{\text{d}}{\text{d}t} x_i = E_{ij} x_j$, with $i,j = 1 \dots 15$. 
When applying an electric field $\boldsymbol{\mathcal{E}}$ the \JMD{wave vector of the Bloch states is modified as} $\hbar \frac{\text{d}}{\text{d}t} \boldsymbol{k} = -e \boldsymbol{\mathcal{E}}$, with $e$ the elementary charge, \JMD{inducing a non-equilibrium component to the spin-pseudospin textures generated by the variation of the magnetic, pseudomagnetic and spin-pseudospin coupling fields in a moving rest-frame.}
\JMD{The spin-pseudospin vector is thus decomposed as $\vec{x} = \vec{x}_\text{eq} + \delta\vec{x}$, where $\vec{x}_\text{eq}$ contains the equilibrium spin-pseudospin texture which can be obtained from the band structure and satisfies $E \,\vec{x}_\text{eq} = 0$. The non-equilibrium component is represented by $\delta \vec{x}$, which is proportional to the applied electric field, and thus within the linear response regime satisfies $\frac{\text{d}}{\text{d}t} \delta\vec{x} \sim \mathcal{O}(\mathcal{E}^2) \approx 0$. The spin-pseudospin dynamic equations now read
\begin{equation}
    -\frac{e}{\hbar} \left( \boldsymbol{\mathcal{E}} \cdot \nabla_{\boldsymbol{k}} \right) \vec{x}_\text{eq} = E \,\delta\vec{x} \text{ .}
    \label{eq:non-eq-x}
\end{equation}
}At this point we must consider that the Ehrenfest matrix cannot by directly inverted in order to solve for the non-equilibrium spin-pseudospin vector. Indeed, if $E$ were invertible then one would obtain an identically null equilibrium spin-pseudospin texture $\vec{x}_\text{eq} = 0$. Since the set of equations represented by $E$ are not linearly independent, then a set of variables $q_i = Q_{ij} \, x_j$ exist such that $E_{ij} \, q_j = \frac{\text{d}}{\text{d}t} q_i = 0$, corresponding to conserved quantities. We thus impose the constrain $\delta q_i = 0$ on Eq.~\eqref{eq:non-eq-x}, by which the matrix problem can be inverted in order to obtain the non-equilibrium spin-pseudospin texture $\delta \vec{x}$. \JMD{We finally note that throughout this procedure disorder dependent processes (such as the momentum relaxation time $\tau$) do not play a role in the emergence of the non-equilibrium spin texture.}

\JMD{\subsection{Validation of Semi-classical Transport Against Kubo-Bastin Formula}
As a means of validating the developed transport methodology, we show that it yields equivalent results to the Kubo-Bastin transport formalism in the weak disorder regime. We consider the Fermi-surface and Fermi-sea decomposition of the Kubo-Bastin formula proposed in Ref.~\cite{bonbien_symmetrized_2020}. The weak disorder limit is represented by a vanishing Kubo-Bastin broadening $\eta \rightarrow 0$, where the Fermi-surface contribution converges to the standard Boltzmann transport contribution setting $\tau = \hbar / \eta$~\cite{bonbien_symmetrized_2020, manchon_current-induced_2019}. On the other hand, the Fermi-sea contribution to the non-equilibrium spin density reads
\begin{equation}
\begin{split}
    \boldsymbol{S}_\text{sea}(\varepsilon_\text{F}) \rightarrow &\, \hbar \int \frac{\text{d}^2 \boldsymbol{k}}{(2\pi)^2} \sum_{n \neq m} \frac{\mathit{f}(E_{n,\boldsymbol{k}}) - \mathit{f}(E_{m,\boldsymbol{k}})}{(E_{n,\boldsymbol{k}} - E_{m,\boldsymbol{k}})^2} \,\text{Im}\, \langle E_{n,\boldsymbol{k}} | \boldsymbol{j} \cdot \boldsymbol{\mathcal{E}} | E_{m,\boldsymbol{k}} \rangle \langle E_{m,\boldsymbol{k}} | \boldsymbol{s} | E_{n,\boldsymbol{k}} \rangle
    \\ = &\, \sum_n \int \frac{\text{d}^2 \boldsymbol{k}}{(2\pi)^2} \mathit{f}(E_{n,\boldsymbol{k}}) \left[ \sum_{m \neq n} 2\hbar \frac{\text{Im}\, \langle E_{n,\boldsymbol{k}} | \boldsymbol{j} \cdot \boldsymbol{\mathcal{E}} | E_{m,\boldsymbol{k}} \rangle \langle E_{m,\boldsymbol{k}} | \boldsymbol{s} | E_{n,\boldsymbol{k}} \rangle}{(E_{n,\boldsymbol{k}} - E_{m,\boldsymbol{k}})^2} \right] \text{ ,}
\end{split}
\label{eq:Ssea_KB}
\end{equation}
with $\boldsymbol{s}$ and $\boldsymbol{j}$ the spin and current density operators respectively. Comparing with Eq.~\eqref{eq:noneqS} we identify that the term within the square brackets in Eq.~\eqref{eq:Ssea_KB} corresponds to the non-equilibrium spin texture $\delta \boldsymbol{s}_{n,\boldsymbol{k}}$, now computed from the Kubo-Bastin formalism. This approach reveals that the non-equilibrium spin texture emerges from inter-band processes related to Berry curvature effects~\cite{li_intraband_2015, dong_berry_2020}. Indeed, this expression for the non-equilibrium spin texture is analogous to that of the anomalous velocity in the context of computing the Hall conductivity Fermi-sea contribution, where the Berry curvature emerges from the non-equilibrium component of the current operator expectation value in $k$-space~\cite{xiao_berry_2010}.

\begin{figure}
    \centering
    \includegraphics[width=\textwidth]{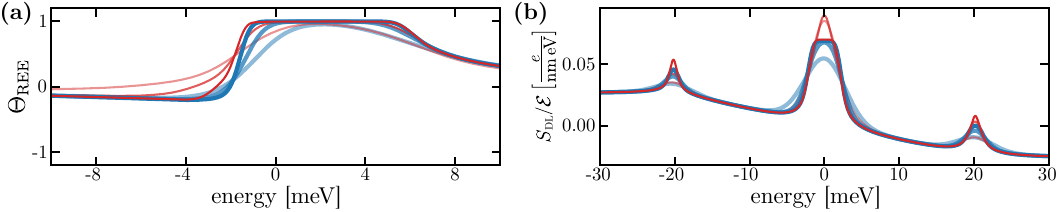}
    \caption{\textbf{(a)} $\Theta_\text{REE}$ and \textbf{(b)} $S_\text{DL}$ transport results computed using both the developed semi-classical spin-dynamics methodology (thick blue curves) and the Kubo-Bastin transport formalism (thin red curves) for broadening values $\eta=2 \,\text{meV}$, $1 \,\text{meV}$ and $0.5 \,\text{meV}$ (lighter to darker curves). The values of other parameters respectively correspond to Figs. 1 and 3 of the main text.}
    \label{fig:SM_methodology}
\end{figure}

We compute the non-equilibrium spin density within the Kubo-Bastin formalism using a $k$-space kernel polynomial expansion method~\cite{medina-duenas_emerging_2024}. In Fig.~\ref{fig:SM_methodology} we show the transport properties contained in Figs.~1 and 3 of the main text; namely, the REE efficiency and the DL non-equilibrium spin density, comparing both methodologies, revealing the convergence of the curves as the pristine limit is approched. Numerical calculations of the Kubo-Bastin formula must always consider a finite broadening parameter $\eta$, which leads to a smearing of the Green's function's singularities. In this context, we additionally introduce a Gaussian energy smearing of variance $\eta^2$ in the semi-classical transport calculations as a means to represent the Kubo-Bastin energy broadening for a better agreement between both methods. We note that this disorder dependent smearing does not modify the semi-classical results beyond merely smearing the sharp features of the curves.}

\section{Spin-Orbit Torque in Graphene: Complete Calculations}
In this section we present detailed calculations and results for the magnetic graphene system with Rashba and Kane-Mele SOC. 

\JMD{We begin calculating the spin texture of the system demonstrating the sharp enhancement or quenching of the spin helicity at the valence and conduction bands}. We treat Rashba SOC which mixes spin up and down subspaces as a perturbation over the non-spin-mixing Hamiltonian, which includes the spin independent term, the magnetic term and Kane-Mele SOC. The unperturbed energies are $E_{\mu\nu} = -\mu J_\text{ex} / 2 + \nu \varepsilon$, with $\mu,\nu=\pm$ and $\varepsilon^2 = (\hbar v k)^2 + \lambda_\text{KM}^2/4$, with $\mu,\nu=\pm$, while the eigenstates $| E_{\mu \nu} \rangle$ correspond to pure states with spin polarization along $\boldsymbol{s}_{\mu \nu}^0 = \mu \hat{\boldsymbol{z}}$ and pseudospin polarization along $\sigma_{\mu \nu}^0 = \nu (\hbar v \boldsymbol{k} - \mu \hat{\boldsymbol{z}} \lambda_\text{KM} / 2 ) / \varepsilon$. At $\hbar v k = \frac{1}{2}(J_\text{ex}^2 - \lambda_\text{KM}^2)^{1/2}$ bands $E_{++}^0$ and $E_{--}^0$ intersect, where Rashba SOC acts non-perturbatively opening a band gap. Indeed, the Rashba Hamiltonian projected to the degenerate subspace is 
\begin{equation}
    \mathcal{H}_{\text{R}, \text{deg}} = \frac{ \lambda_\text{R} \hbar v k}{2 \varepsilon} \begin{pmatrix}
    0 & \text{i} \text{e}^{-\text{i} \varphi} \\
    -\text{i} \text{e}^{\text{i} \varphi} & 0
    \end{pmatrix} \text{ ,}
\end{equation}
opening a band gap of magnitude $\Delta E_\text{g} = |\lambda_\text{R}| [1 - \lambda_\text{KM}^2 / J_\text{ex}^2]^{1/2}$. Remarkably, the hybridization between both bands generates an in-plane spin texture since $\langle E_{++}^0 | \boldsymbol{s} | E_{--}^0 \rangle = \langle +\hat{\boldsymbol{z}} | \boldsymbol{s} | -\hat{\boldsymbol{z}} \rangle \langle \sigma_{++}^0 | \sigma_{--}^0 \rangle = (\hat{\boldsymbol{x}} - \text{i} \hat{\boldsymbol{y}}) \lambda_\text{KM} / 2$. 
The resulting energies are $E_{\text{deg},\nu} = \nu \varepsilon_\text{g}$, with $\varepsilon_\text{g}^2 = (J_\text{ex} / 2 - \varepsilon)^2 + (\lambda_\text{R} \hbar v k / 2\varepsilon)^2$ and $\nu=\pm$ indicating the conduction and valence band respectively. The spin texture is
\begin{equation}
    \boldsymbol{s}_{\text{deg},\nu} = - \nu \frac{J_\text{ex} / 2 - \varepsilon}{\varepsilon_\text{g}} \hat{\boldsymbol{z}} - \nu \frac{\lambda_\text{R} \lambda_\text{KM} \hbar v k}{4 \varepsilon^2 \varepsilon_\text{g}} \hat{\boldsymbol{\varphi}} \text{ ,}
\end{equation}
where we observe that the spin helicity exhibits a sharp peak near the gap marked by the direct gap $2 \varepsilon_\text{g}$, in the denominator of $\boldsymbol{s}_{\text{deg},\nu} \cdot \hat{\boldsymbol{\varphi}}$. The sharp peak of the spin helicity is of opposite sign in the valence and conduction bands, while far from the gap the bands remain spin-polarized along the out-of-plane axis. This peak is generated because bands $E_{++}$ and $E_{--}$ share a common pseudospin subspace, reflected by a non-zero value of the internal product $\langle \sigma_{++}^0 | \sigma_{--}^0 \rangle$.

We remain with a set of four non-degenerate bands ordered from lowest to highest energy as $\{ E_{+,-} ,\, E_{\text{deg},-} ,\, E_{\text{deg},+} ,\, E_{-,+} \}$. Via standard perturbation theory, we compute the spin texture to first order with respect to Rashba SOC, yielding
\JMD{\begin{subequations}
\begin{equation}
    \boldsymbol{s}_{\mu,-\mu} = \mu \hat{\boldsymbol{z}} + \frac{\lambda_\text{R}}{2\varepsilon} \sqrt{1 - \frac{\lambda_\text{KM}^2}{4\varepsilon^2}} \left( \frac{\varepsilon - \mu \frac{1}{2}\lambda_\text{KM}}{\frac{1}{2} J_\text{deg}} + \mu \frac{\lambda_\text{KM}}{\frac{1}{2} J_\text{ex} + \varepsilon} \right) \hat{\boldsymbol{\varphi}} \text{ ,}
\end{equation}
\begin{equation}
    \boldsymbol{s}_{\text{deg},\nu} = -\nu \frac{\frac{1}{2} J_\text{ex} - \varepsilon}{\varepsilon_\text{g}} \hat{\boldsymbol{z}} - \frac{\lambda_\text{R}}{2 \varepsilon} \sqrt{1 - \frac{\lambda_\text{KM}^2}{4 \varepsilon^2}} \left( \frac{\varepsilon - \nu\xi \frac{1}{2} \lambda_\text{KM}}{\frac{1}{2} J_\text{ex}} + \nu \frac{\frac{1}{2}\lambda_\text{KM}}{\varepsilon_\text{g}} \right) \hat{\boldsymbol{\varphi}} \text{ ,}
\end{equation}
\end{subequations}
with $\chi = \text{sign}\, (\frac{1}{2} J_\text{ex} - \varepsilon)$. We note that even though the spectrum to first order in $\lambda_\text{R}$ remains symmetric, as $E_{\mu,-\mu} = -E_{-\mu,\mu}$ and $E_{\text{deg},\nu} = -E_{\text{deg},-\nu}$, broken particle-hole symmetry is manifest because the spin texture of the symmetric bands cannot be related by a unitary transformation if $\lambda_\text{KM} \neq 0$.}

\begin{figure}
    \centering
    \includegraphics[width=\textwidth]{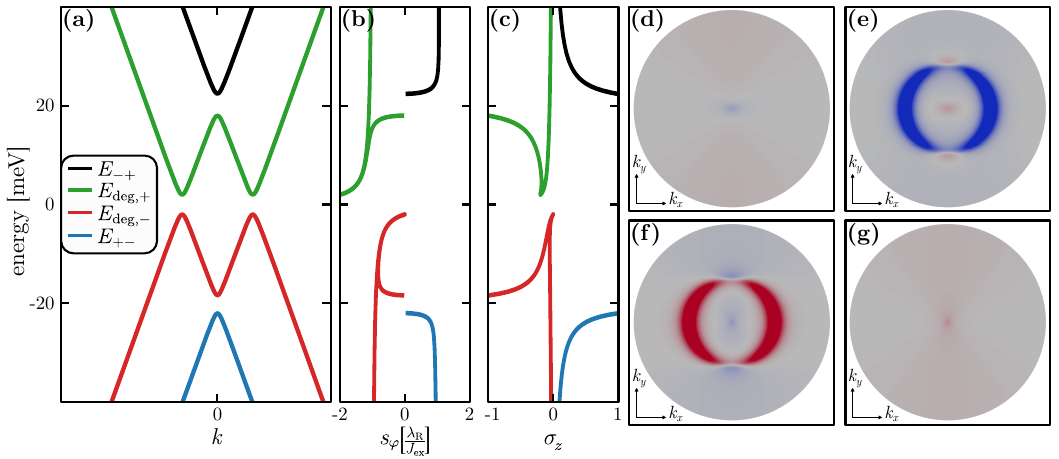}
    \caption{With Kane-Mele SOC ($\lambda_\text{KM} = \lambda_\text{R} = 4 \,\text{meV}$, and $J_\text{ex}=40 \,\text{meV}$): \textbf{(a)} Band structure, \textbf{(b)} helical spin texture and \textbf{(c)} out-of-plane pseudospin texture, while panels \textbf{(d)--(g)} show the damping-like non-equilibrium spin texture $\delta s_\text{DL}$ of the bands in decreasing energy order.}
    \label{fig:SM_neqS1}
\end{figure}

\begin{figure}
    \centering
    \includegraphics[width=\textwidth]{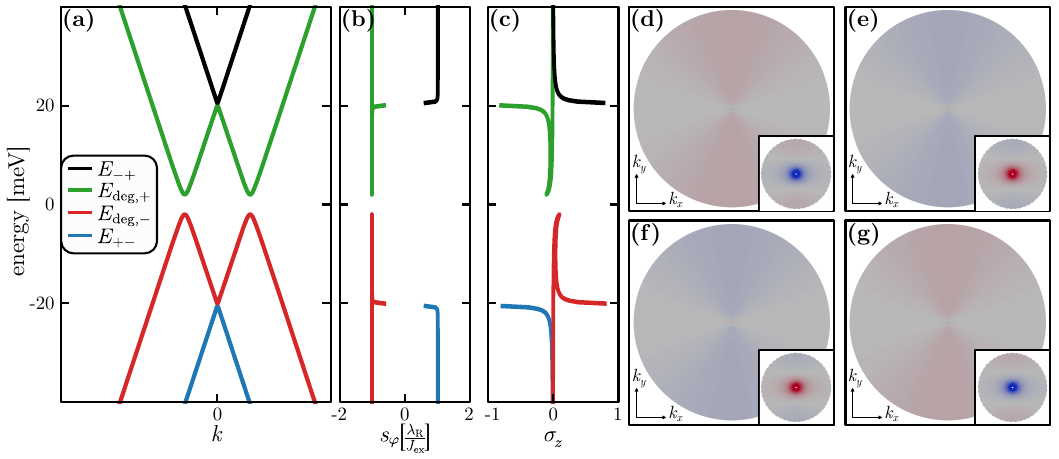}
    \caption{Same as Fig. \ref{fig:SM_neqS1}, without Kane-Mele SOC ($\lambda_\text{KM} = 0$). The $\delta s_\text{DL}$ scale in panels (d)--(g) is twice in comparison to Fig. \ref{fig:SM_neqS1}, while the inset shows a zoom to $k=0$.}
    \label{fig:SM_neqS0}
\end{figure}

We have shown how Kane-Mele SOC breaks particle-hole symmetry in the system and generates a sharp variation of the spin helicity near the band gap.
Figs. \ref{fig:SM_neqS1} and \ref{fig:SM_neqS0} show the band structure with the relevant spin and pseudospin texture components, and the damping-like non-equilibrium spin density $\delta s_\text{DL}$ for each band, with and without Kane-Mele SOC respectively. 
\JMD{In Fig.~\ref{fig:SM_neqS1}(c) we observe that for $\lambda_\text{KM} \neq 0$ bands $E_{\text{deg},-}$ and $E_{\text{deg},+}$ share a common pseudospin component when approaching the band gap, which enables the sharp variation of the spin helicity, shown in panel (b), where the spin helicity of the valence band is quenched while that of the conduction band is enhanced. On the other hand, in absence of Kane-Mele SOC (see Fig.~\ref{fig:SM_neqS0}), the pseudospin polarization of these bands are exactly opposite and thus the spin helicity remains constant near the gap.}
\JMD{Panels (d)--(g) show the damping-like non-equilibrium spin texture generated within each band, where it is clear that the sharp variation of the spin helicity near the band inversion generates a large source for the damping-like torque. Additionally, we observe an $S_\text{DL}$ source at the Dirac points. When comparing the $\lambda_\text{KM} = 0$ and $\lambda_\text{KM} = |\lambda_\text{R}|$ systems, respectively shown in Figs.~\ref{fig:SM_neqS0} and \ref{fig:SM_neqS1}, we observe that the $S_\text{DL}$ source at the Dirac point changes sign for bands $E_{+-}$ and $E_{\text{deg},-}$, while remains of the same sign in bands $E_{-+}$ and $E_{\text{deg},+}$. As explained in the main text, this occurs because Rashba SOC induces an effective mass of $\sim \lambda_\text{R}^2 / J_\text{ex}$ on both spin-split Dirac cones, while Kane-Mele SOC induces an effective mass of opposite sign in each cone.}

\begin{figure}[t]
    \centering
    \includegraphics[width=\textwidth]{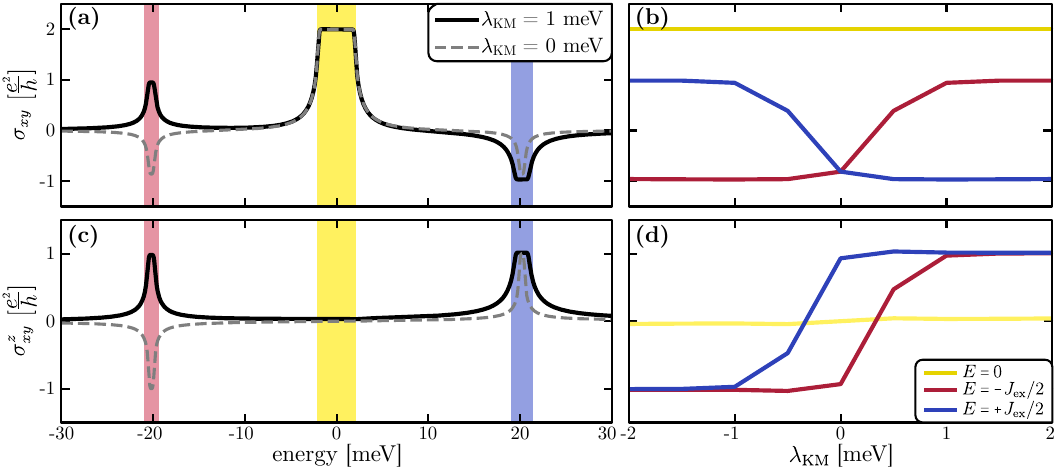}
    \caption{\JMD{Hall ($\sigma_{xy}$) and spin Hall ($\sigma_{xy}^z$) conductivity in the magnetic Rashba graphene system, for different values of Kane-Mele SOC ($\lambda_\text{KM}$), with $J_\text{ex} = 40 \,\text{meV}$ and $\lambda_\text{R} = 4 \,\text{meV}$}.}
    \label{fig:SM_Hall}
\end{figure}

\JMD{Our SOT results indicate the emergence of $k$-space sources of damping-like torque along with the emergence of topological edge states. Indeed, a quantum anomalous Hall phase has been proposed in magnetic Rashba graphene at the band gap at charge neutrality~\cite{qiao_quantum_2010}, while the Dirac point semi-gap is opened by Kane-Mele SOC, which is known to generate a spin Hall current due to Berry curvature effects~\cite{kane_z_2005}. Fig.~\ref{fig:SM_Hall} shows the Hall and spin Hall conductivity in the system. We observe a quantized anomalous Hall conductivity of $\sigma_{xy} = 2 e^2/h$ throughout the band gap at charge neutrality, as seen in the yellow highlight in Fig.~\ref{fig:SM_Hall}(a), which we find is preserved when including Kane-Mele SOC, as shown in panel (b). For $\lambda_\text{KM} = 0$ the quantized Hall conductivity is spinless; however, when including Kane-Mele SOC a non-zero spin Hall conductivity emerges, as shown by panels (c) and (d). Indeed, for $\lambda_\text{KM} \neq 0$ the quantum anomalous Hall edge states must continuously bridge the difference in the spin texture between the valence and conduction bands, thus acquiring distinct spin properties resulting in a non-zero spin Hall current. At the spin-split Dirac points, highlighted in red and blue, we observe that even in the absence of Kane-Mele SOC, a nearly quantized Hall conductivity of $|\sigma_{xy}| = e^2/h$ which is almost fully spin polarized emerges, indicating that the Hall current is generated within the semi-gap of the spin-polarized Dirac cones. When including Kane-Mele SOC, with $\lambda_\text{KM} > 0$, we find that the Hall and spin Hall conductivity reverse at the spin majority Dirac point, and maintain their sign in the spin minority cone, consistent with the damping-like torque behavior.}

\JMD{We finally discuss the effects of other interactions, such as valley-Zeeman SOC and a staggered sublattice potential, which may emerge from broken sublattice symmetry due to the adjacent ferromagnetic layer. Valley-Zeeman SOC enters the Hamiltonian as a term proportional to $\tau_z s_z$, which results in a valley-dependent exchange field. This lifts the valley degeneracy; however, the emerging physics is adequately captured by our analysis simply by separately analyzing each valley and adjusting the exchange parameter $J_\text{ex}$. The staggered sublattice potential is represented by a term proportional to $\sigma_z$ in the original Hamiltonian, which after transforming the $\tau=-$ subspace results in an effective mass of opposite sign on each valley. This term also lifts the valley degeneracy, and the emerging physics may be captured by adjusting the effective mass induced in each Dirac cone, considering all contributions emerging from Rashba SOC, Kane-Mele SOC, and the staggered sublattice potential.}